\documentclass[11pt]{prop_axis} % MM proposal style

\usepackage[super,comma]{natbib} % bibliography
\usepackage{apjfonts} %use times font for math
\usepackage{graphicx} % Graphics
\usepackage{amssymb,amsmath} % Math/Greek symbols
\usepackage{color} % colorize text
\usepackage{enumitem} % better lists
\usepackage{wrapfig} % Wrap text around figures
\usepackage[table]{xcolor} % table row colors
\usepackage{multirow} % combine multiple rows in a table
\usepackage{tabularx} % better tables
\usepackage{array} % for tables

\usepackage{hyperref} % links within the document
\hypersetup{
    colorlinks,
    linkcolor={blue!80!black},
    citecolor={blue!80!black},
    urlcolor={blue!80!black}
}

\newcolumntype{L}[1]{>{\raggedright\let\newline\\\arraybackslash\hspace{0pt}}p{#1}}
\newcolumntype{C}[1]{>{\centering\let\newline\\\arraybackslash\hspace{0pt}}p{#1}}
\newcolumntype{R}[1]{>{\raggedleft\let\newline\\\arraybackslash\hspace{0pt}}p{#1}}
\renewcommand\arraystretch{1.3}

\definecolor{headcolor}{rgb}{0.65,0.65,0.65}
\usepackage{fancyhdr}
\renewcommand{\topmargin}{-9mm}

\renewcommand{\textfloatsep}{4mm}

\newcommand{\lem}{\textit{LEM}}
\newcommand{\LEM}{\textit{LEM}}

\newcommand{\fermi}{\textit{Fermi}}

\newcommand{\chandra}{\textit{Chandra}}
\newcommand{\Chandra}{\textit{Chandra}}
\newcommand{\xmm}{\textit{XMM-Newton}}
\newcommand{\XMM}{\textit{XMM-Newton}}
\newcommand{\lynx}{\textit{Lynx}}

\newcommand{\hubs}{\textit{HUBS}}

\newcommand{\Athena}{\textit{Athena}}
\newcommand{\athena}{\textit{Athena}}

\newcommand{\erosita}{\textit{eROSITA}}
\newcommand{\XRISM}{\textit{XRISM}}
\newcommand{\xrism}{\textit{XRISM}}
\newcommand{\hitomi}{\textit{Hitomi}}

\newcommand{\Hubble}{\textit{Hubble}}

\newcommand{\SKA}{\textit{SKA}}

\newcommand{\halosat}{\textit{HaloSat}}

\newcommand{\as}{$^{\prime\prime}$}

\newcommand{\am}{$^{\prime}$}

\newcommand{\msun}{~$M_{\odot}$}

\def\gax{\gtrsim}

\newcommand{\bsf}{\sffamily\bfseries}
\newcommand{\bs}{\sffamily\bfseries\small}
\newcommand{\bi}{\itshape\bfseries}
\newcommand{\sfsm}{\sffamily\small}
\newcommand{\smsk}{\vspace{1mm}}

\definecolor{callout}{rgb}{0.25,0.40,0.85}
\definecolor{calllem}{rgb}{0.20,0.45,0.99}
\definecolor{tabledef}{rgb}{0.95,0.95,0.95}
\definecolor{tablealt}{rgb}{0.77,0.80,1.0}
\definecolor{tablelem}{rgb}{0.80,0.85,1.0}
\definecolor{whitelem}{rgb}{1.0,1.0,1.0}
\definecolor{greenlem}{rgb}{0.7,1.0,0.7}

\begin{document}

% this is single-space for 11pt font: 11*1.2=13.2
\baselineskip=13.2pt
% (and baselinestretch is 1.0)
\sloppy
\pagenumbering{roman}
\thispagestyle{empty}
%\pagecolor{black}

%%%%%%%%%%%%%%%%%%%%%%%%%%%%%%%%%%%%%%%%%%%%%%%%%%%%%%%%%%%%%%%
%%% cover page
\begin{figure}[t]
%\vspace*{15mm}
%\centering
\vspace*{-25mm}
\hspace*{-26mm}
\includegraphics[width=8.5in]{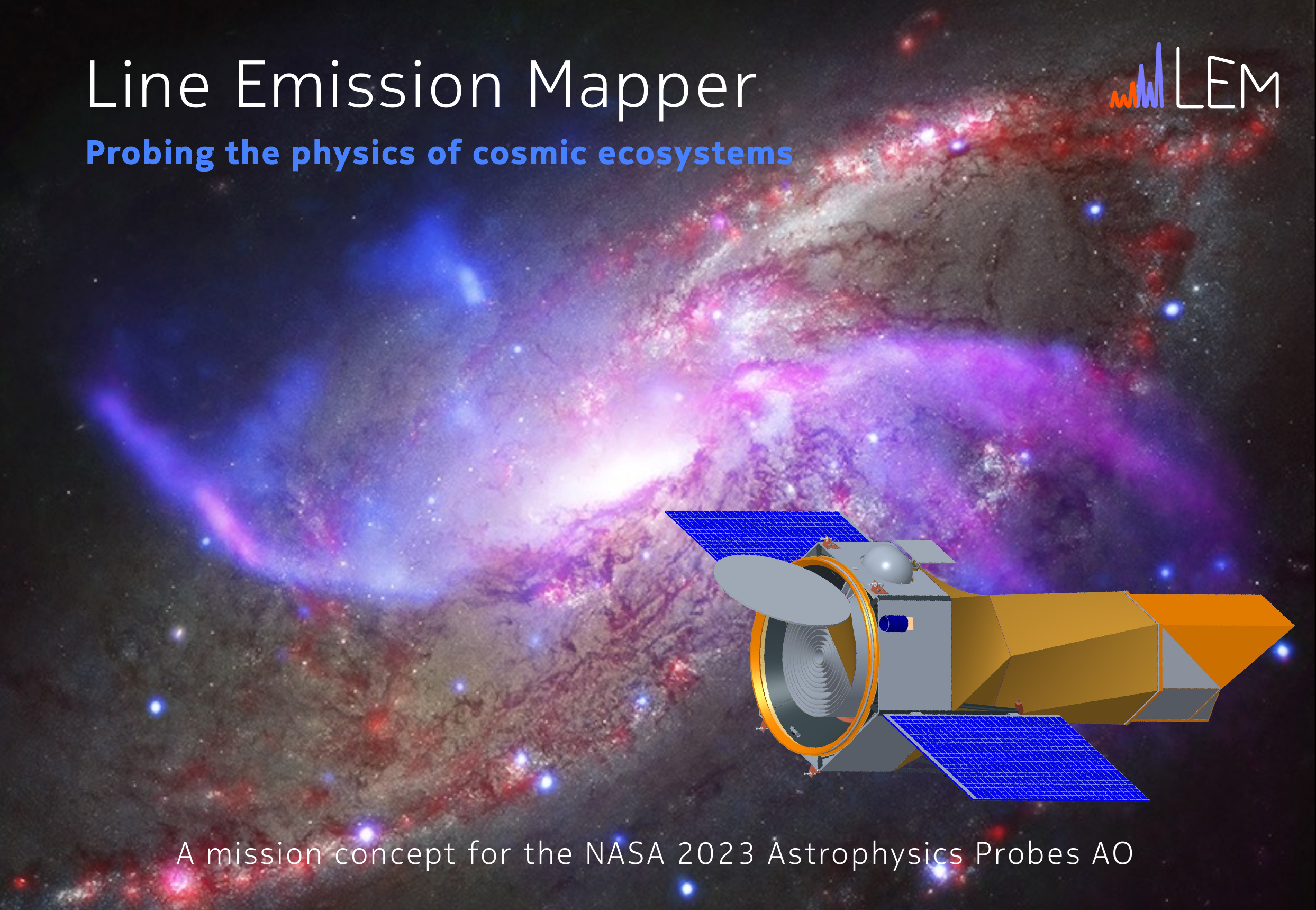}

\label{fig:cad_ovlay}
\end{figure}
%%%%%%%%%%%%%%%%%%%%%%%%%%%%%%%%%%%%%%%%%%%%%%%%%%%%%%%%%%%%%%%

%%%%%%%%%%%%%%%%%%%%%%%%%%%%%%%%%%%%%%%%%%%%%%%%%%%%%%%%%%%%
% \title{\textcolor{black}{\sf\huge THE LINE EMISSION MAPPER}}
% \maketitle

% \vspace*{2mm}
% \centerline{\textcolor{headcolor}{\sffamily\bfseries\large
% A MISSION CONCEPT FOR THE NASA 2023 ASTROPHYSICS PROBE AO}}

% authors: (for some reason, \author{} doesn't work)
\vspace*{-11mm}
\begin{center}
\begin{minipage}{17.5cm}
\hspace{-5mm}
\centering
Ralph~Kraft$^{1}$,
Maxim~Markevitch$^{2}$,
Caroline~Kilbourne$^{2}$,
Joseph~S.~Adams$^{2}$,
Hiroki~Akamatsu$^{3}$,
Mohammadreza~Ayromlou$^{4}$,
Simon~R.~Bandler$^{2}$,
Marco~Barbera$^{51,52}$,
Douglas~A.~Bennett$^{5}$,
Anil~Bhardwaj$^{6}$
% ...
\end{minipage}
\end{center}

\vfill
\centerline{\em Draft mission concept, March 2023}
\clearpage
\pagecolor{white}

\twocolumn

{\footnotesize
\noindent
$^{1}$~~Center for Astrophysics $|$ Harvard \& Smithsonian\\
$^{2}$~~NASA Goddard Space Flight Center\\
$^{3}$~~SRON, the Netherlands\\
$^{4}$~~Universit{\"a}t Heidelberg, Germany\\
$^{5}$~~NIST, Boulder, Colorado\\
$^{6}$~~Physical Research Laboratory, Ahmedabad, India\\
% ...
$^{51}$~Universit{\`a} degli Studi di Palermo, Italy\\
$^{52}$~INAF, Palermo, Italy\\
% ...
\\
\phantom{${^52}$}~\textcolor{blue}{\bsf lem-observatory.org}\\
}

% %%%%%%%%%%%%%%%%%%%%%%%%%%%%%%%%%%%%%%%%%%%%%%%%%%%%%%%%%%%%%%%
% \begin{figure}[b]
% %\vspace*{15mm}
% \includegraphics[width=8cm]{figures/3logos_white.pdf}
% 
% %\vspace*{-10mm}
% \label{fig:logos}
% \end{figure}
% %%%%%%%%%%%%%%%%%%%%%%%%%%%%%%%%%%%%%%%%%%%%%%%%%%%%%%%%%%%%%%%

\clearpage
\twocolumn

% \renewcommand{\contentsname}{ }
% {\sf\small\vspace*{-17mm}
% \tableofcontents
% }
% \clearpage

\setcounter{page}{1}
\pagenumbering{arabic}

\section{SUMMARY}
\label{sec:summary}

The Line Emission Mapper (\lem) is an X-ray Probe for the 2030s that will
answer the outstanding questions of the Universe's structure formation. It
will also provide transformative new observing capabilities for every area
of astrophysics, and to heliophysics and planetary physics as well. \lem's
main goal is a comprehensive look at the physics of galaxy formation,
including stellar and black-hole feedback and flows of baryonic matter into
and out of galaxies. These processes are best studied in X-rays; as
emphasized by the 2020 Decadal Survey, emission-line mapping is the pressing
need in this area. \lem\ will use a large microcalorimeter array/IFU (that
builds on \athena\ XIFU technology developments), covering a
$30\times30$\am\ field with 10\as\ angular resolution, to map the soft X-ray
line emission from objects that constitute galactic ecosystems. These
include supernova remnants, star-forming regions, superbubbles, galactic
outflows (such as the Fermi/eROSITA bubbles in the Milky Way and their
analogs in other galaxies), the Circumgalactic Medium in the Milky Way and
other galaxies, and the Intergalactic Medium at the outskirts and beyond the
confines of galaxies and clusters. \lem's 1--2 eV spectral resolution in the
0.2--2 keV band will make it possible to disentangle the faintest emission
lines in those objects from the bright Milky Way foreground, providing
groundbreaking measurements of the physics of these plasmas, from
temperatures, densities, chemical composition to gas dynamics. While the
mission is optimized to provide critical observations that will push our
understanding of galaxy formation, \LEM\ will provide transformative
capability for all classes of astrophysical objects, from the Earth's
magnetosphere, planets and comets to the interstellar medium and X-ray
binaries in nearby galaxies, AGN, and cooling gas in galaxy clusters. In
addition to pointed observations, \lem\ will perform a shallow all-sky
survey that will dramatically expand the discovery space.

\section{MISSION CONCEPT}
\label{sec:concept}

\begin{figure}[tb]
\centering
%\vspace*{-5mm}
\includegraphics[width=0.4\textwidth]{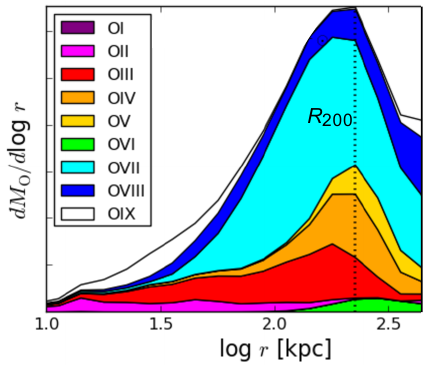}

%\vspace*{2mm}
\caption{Why soft X-rays? Predicted fraction of Oxygen ions in a CGM halo of
a Milky-Way mass galaxy (simulations using EAGLE model\cite{Oppenheimer16};
dotted line is the virial radius; vertical scale is linear). Probing this
circumgalactic gas is critical to understanding the physics of galaxies. The
temperature of the halo is such that the dominant species over most of the
halo is O{\sc vii}, whose emission (along with that of O{\sc viii}) lies in
the soft X-ray band --- the energy of the Oxygen He$\,\alpha$ resonant line
is 574 eV. The lower ionization species are observable in the UV, but
modeling and extrapolation are required to deduce the properties of the
CGM. An X-ray instrument will probe the bulk of the CGM directly,
complementing the UV data on the cooler gas phases.}
\vspace*{1mm}

\label{fig:ofrac}
%\end{wrapfigure}
\end{figure}

The Line Emission Mapper (\lem) is an X-ray Probe concept aimed at one of
the fundamental unsolved problems of modern astrophysics --- galaxy
formation. Galaxies are governed by the competition of gravity and the
energetic ``feedback'' from the forming stars and black holes. Theoretical
studies predict that key information on these processes is encoded in the
tenuous Circumgalactic Medium (CGM) and even more rarefied Intergalactic
Medium (IGM), which together contain most of the baryonic matter in the
Universe\cite{Cen99}. Most of that matter has so far eluded detection, much
less detailed study. The Astro2020 Decadal Survey has identified mapping CGM
and IGM {\em in emission}\/ as the key missing observing capability and a
major ``discovery area.'' \lem\ is designed to provide this new capability
in the soft X-ray band, where the dominant CGM and IGM ion species emit
(Fig.\ \ref{fig:ofrac}). While the importance of studying these hidden
baryons has been recognized for decades, only now has the technology arrived
that makes it possible.

%%%%%%%%%%%%%%%%%%%%%%%%%%%%%%%%%%%%%%%%%%%%%%%%%%%%%%%%%%%%%%%%%%%%%%%%%%
%\begin{wrapfigure}{r}{0.49\textwidth}
%%%%%%%%%%%%%%%%%%%%%%%%%%%%%%%%%%%%%%%%%%%%%%%%%%%%%%%%%%%%%%%

%%%%%%%%%%%%%%%%%%%%%%%%%%%%%%%%%%%%%%%%%%%%%%%%%%%%%%%%%%%%%%%
\begin{figure*}[t]
\centerline{%
\includegraphics[width=13cm]{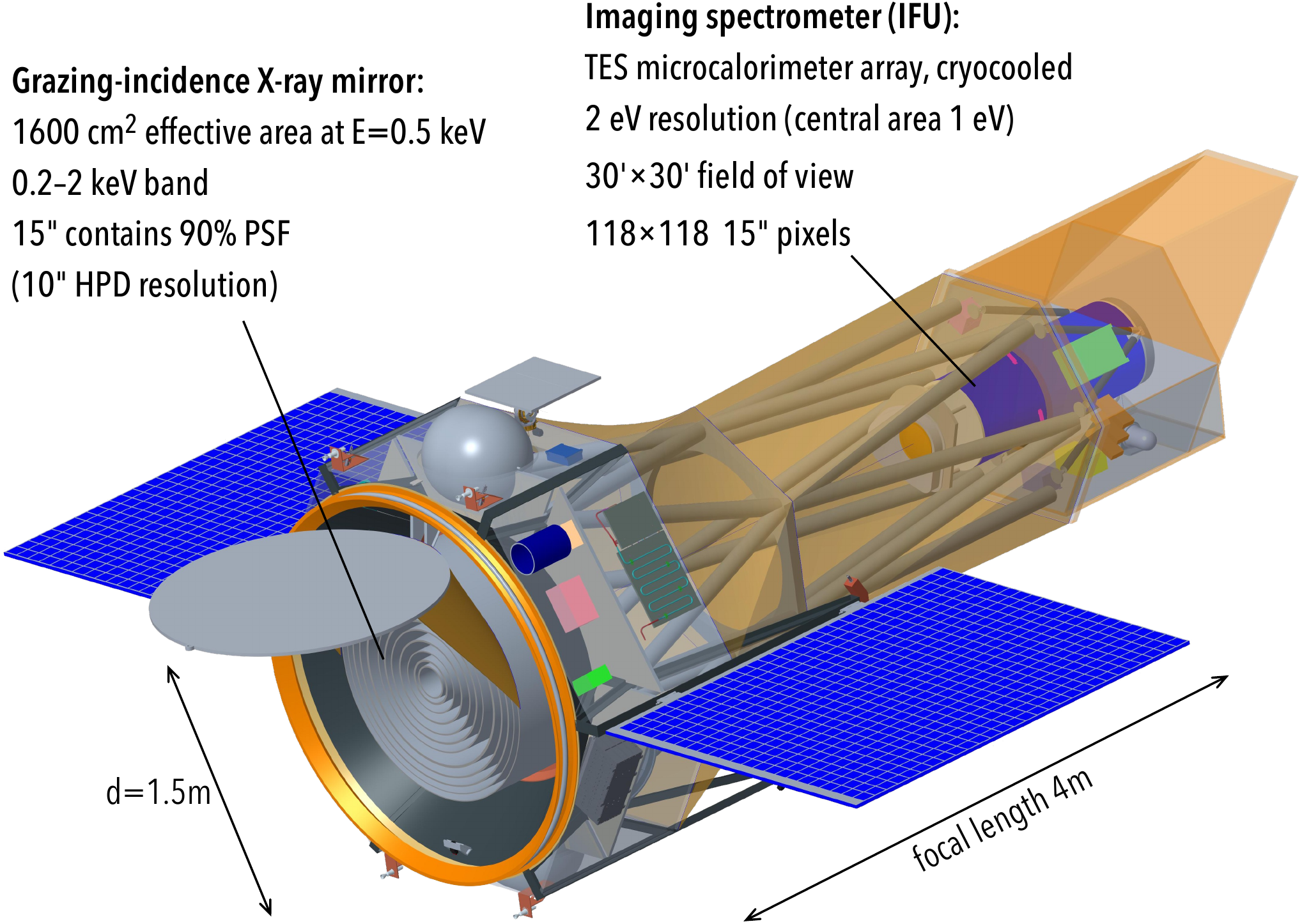}
}

\vspace*{3mm}
\caption{\LEM\ mission design (under development in collaboration with
Lockheed Martin Corp.; a snapshot as of November 2022 is shown).}

\label{fig:crosssec}
\end{figure*}
%%%%%%%%%%%%%%%%%%%%%%%%%%%%%%%%%%%%%%%%%%%%%%%%%%%%%%%%%%%%%%%

%%%%%%%%%%%%%%%%%%%%%%%%%%%%%%%%%%%%%%%%%%%%%%%%%%%%%%%%%%%%%%%
%\begin{wrapfigure}{r}{0.8\textwidth}
\begin{figure*}[tbp]
\vspace*{5mm}
\centerline{%
\includegraphics[width=7cm]{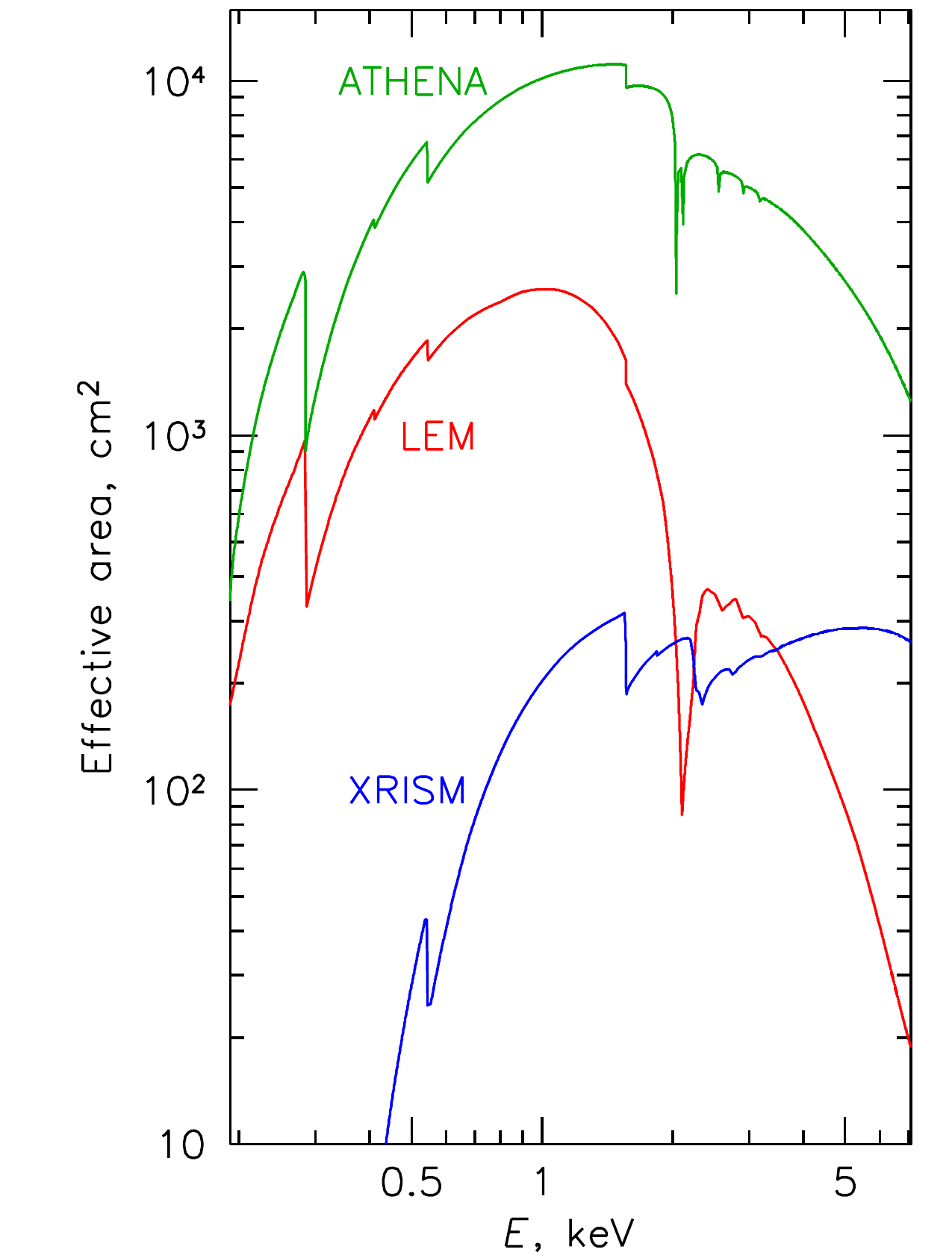}
\includegraphics[width=7cm]{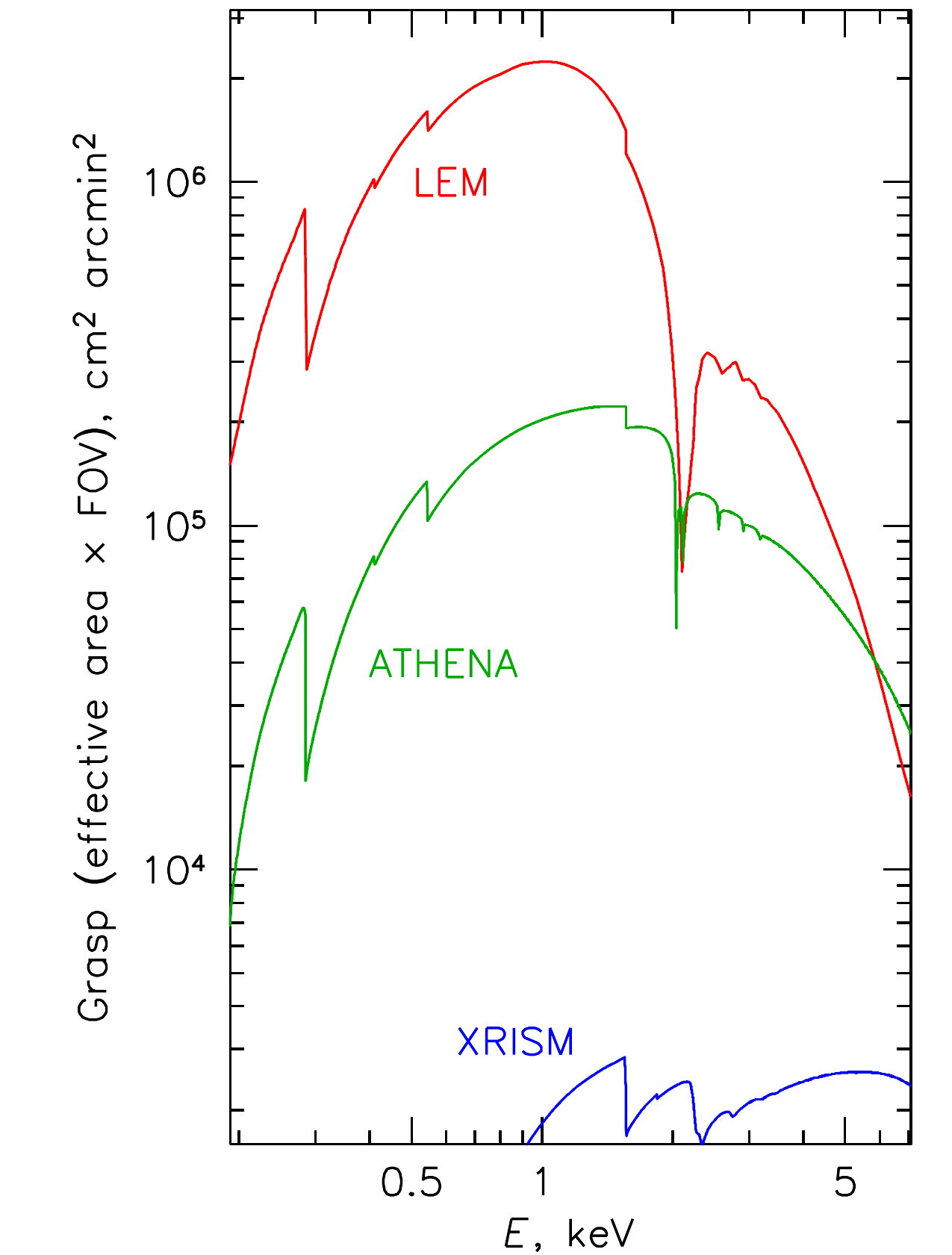}
}

\vspace*{1mm}
\caption{\LEM\ effective area and grasp --- the product of the effective
area and the field of view solid angle --- compared to \XRISM\ Resolve and
\Athena\ XIFU (prior to 2022 reformulation). Grasp is the quantity relevant
for mapping faint extended objects with sizes comparable to or greater than
the field of view (such as nearby galaxies, clusters, the IGM, or the Milky
Way structures) --- for such targets, the number of photons collected in a
given exposure is directly proportional to grasp.}
\vspace*{5mm}

\label{fig:areas}
\end{figure*}
%\end{wrapfigure}
%%%%%%%%%%%%%%%%%%%%%%%%%%%%%%%%%%%%%%%%%%%%%%%%%%%%%%%%%%%%%%%

\lem\ is a single telescope that consists of an X-ray mirror with a large
collecting area and moderate angular resolution, and a cryogenic
microcalorimeter array (Fig.\ \ref{fig:crosssec}). Both are optimized for
the 0.2--2 keV energy band, where the most abundant CGM/IGM ion species have
multiple bright emission lines. Because the CGM and IGM are diffuse sources,
mapping them requires a {\em nondispersive}\/ imaging spectrometer, or an
integral field unit (IFU) as they are known in other wavebands.
(\chandra\ HETG and LETG, \xmm\ RGS and the {\em ARCUS}\/ concept are
dispersive spectrometers whose full spectral resolution is available only
for point sources.)  Furthermore, as will be seen below, our own Milky Way
places a bright screen in front of the very faint IGM and CGM around other
galaxies. Removing this foreground requires eV-class spectral
resolution. These two conditions require a microcalorimeter. Such imaging
X-ray calorimeters have not been flown before, with the exception of the
short-lived \hitomi, to be replaced in 2023 by \xrism. \lem\ will provide a
leap beyond \xrism, and surpass in relevant capabilities other missions
planned for the more distant future. Table~\ref{table:mission_comp} compares
\lem\ to future X-ray microcalorimeters --- namely, \xrism\ Resolve,
\athena\ XIFU (to be launched in late 2030s) and \hubs\ (under study). The
effective area and ``grasp'' as functions of energy for \lem, \xrism\ and
\athena\ are compared in Fig.\ \ref{fig:areas}. We note that \athena\ is
currently undergoing redefinition and as of this writing its final
instrument parameters are unknown. The above comparisons are done with the
pre-2022 \athena\ concept.

The \lem\ concept is being prepared for the NASA 2023 Astrophysics Probes
call for proposals. If selected, \lem\ will be launched in 2032.

%%%%%%%%%%%%%%%%%%%%%%%%%%%%%%%%%%%%%%%%%%%%%%%%%%%%%%%%%%%%%%%
{\rowcolors{3}{tabledef}{tablealt}
\begin{table*}[tbp]
\centering\small
\vspace*{5mm}
\begin{tabularx}{1.0\textwidth}{ L{5.cm} C{2.9cm} C{2.5cm} C{2.35cm} C{1.7cm} }
%\hline
\rowcolor{callout}
\phantom{\raisebox{-3mm}{\rule{0cm}{8mm}}}&
\cellcolor{calllem}\textcolor{white}{\sfsm LEM} &
\textcolor{white}{\sfsm XRISM Resolve} &
\textcolor{white}{\sfsm Athena XIFU$^*$} &
\textcolor{white}{\sfsm HUBS} \\
%\hline
Energy band, keV & \cellcolor{tablelem}0.2--2 & 0.4--12 & 0.2--12 & 0.2--2\\
Effective area, cm$^2$%
\hfill 0.5 keV \newline \hfill 6 keV &
\cellcolor{whitelem}1600\newline 0 &
50\newline 300 &
6000\newline 2000 &
500\newline 0\\
Field of view &\cellcolor{tablelem}30\am\ & 3\am\ & 5\am\ & 60\am \\
Grasp, $10^4$ cm$^2$ arcmin$^2$ \hfill 0.5 keV &
\cellcolor{greenlem}140 & 0.05 &12 & 180\\
Angular resolution &\cellcolor{tablelem}15\as\ & 75\as\ & 5\as\ & 60\as\\
Spectral resolution &\cellcolor{greenlem}1 eV (central 7\am), \newline 2
eV (rest of FOV) & 7 eV & 2.5 eV & 2 eV\\
Detector size, pixels (equiv.\ square) &\cellcolor{tablelem}118$\times$118 & 6$\times$6 & 50$\times$50 & 60$\times$60 \\
%\hline
\end{tabularx}
\vspace*{3mm}
\caption{\LEM\ compared with the requirements for the future X-ray
nondispersive spectrometers onboard \XRISM\ (JAXA/NASA), \Athena\ (ESA/NASA)
and \hubs\ (Chinese Space Agency).  For \xrism, based on \hitomi, the actual
spectral resolution is expected to be 5 eV. The main advantages of \lem\ are
its large grasp and high spectral resolution. $^*$For \athena,
pre-reformulation requirements are listed.}
\label{table:mission_comp}
%\vspace*{5mm}
\end{table*}
}
%%%%%%%%%%%%%%%%%%%%%%%%%%%%%%%%%%%%%%%%%%%%%%%%%%%%%%%%%%%%%%%

%%%%%%%%%%%%%%%%%%%%%%%%%%%%%%%%%%%%%%%%%%%%%%%%%%%%%%%%%%%%%%%
\begin{figure}[b]
\centering
\vspace*{3mm}
\includegraphics[width=0.38\textwidth]{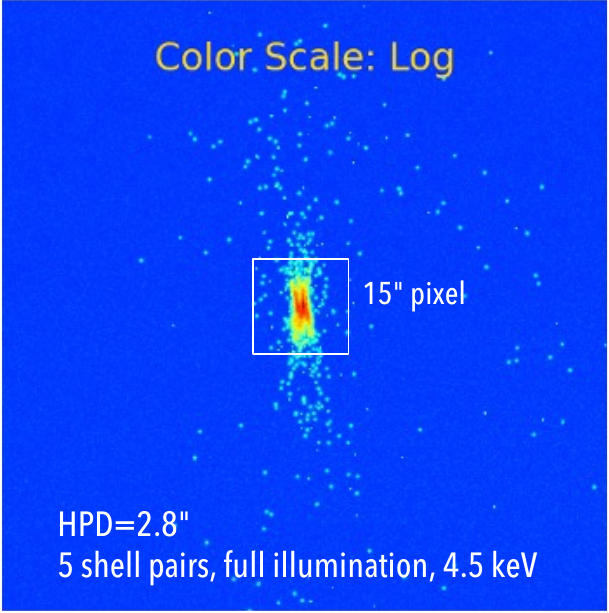}

\vspace{1.8mm}
\caption{Point Spread Function measurement for a subset of the mirror
consisting of 5 pairs of reflective shells. The image is elongated because
it comes from an incomplete mirror. A 2.8\as\ HPD is achieved for this 
subset; the requirement is 10\as\ HPD (or, equivalently, for 90\% of the
photons to fall in the 15\as\ detector pixel) for the whole \lem\ mirror
that will combine approximately 3500 shell pairs.}

\label{fig:psf}
\end{figure}
%%%%%%%%%%%%%%%%%%%%%%%%%%%%%%%%%%%%%%%%%%%%%%%%%%%%%%%%%%%%%%%

%%%%%%%%%%%%%%%%%%%%%%%%%%%%%%%%%%%%%%%%%%%%%%%%%%%%%%%%%%%%%%%
\begin{figure*}
\centering
%\vspace*{3.5mm}
\includegraphics[width=0.95\textwidth]{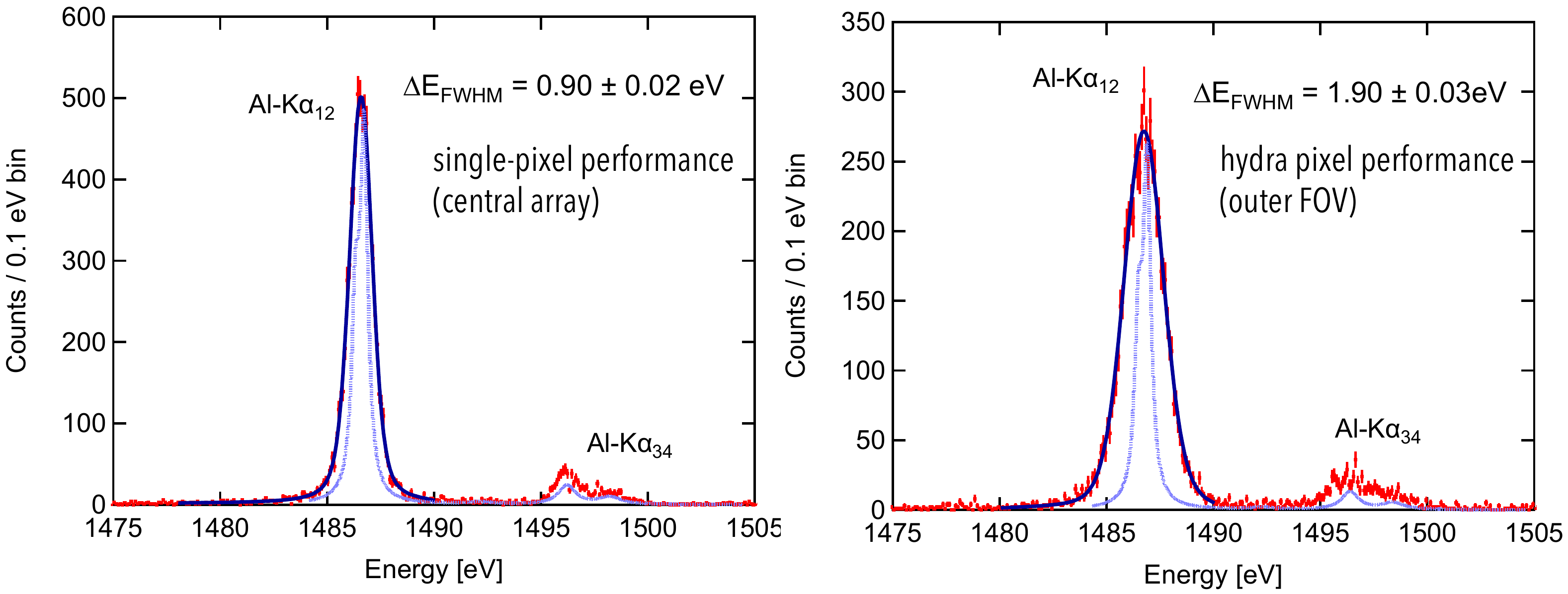}

%\vspace{1.8mm}
\caption{Spectral resolution for a prototype \lem\ detector
array\cite{Smith23}, measured for the single-absorber pixels (the central
array) and the 4-pixel hydras (the rest of the detector). The measurements
were performed at $E=1.5$ keV. The blue dotted line is the intrinsic Al line
profile (arbitrary normalization) and the black line is the model that includes
detector broadening. The performance already nearly meets the
\lem\ requirements.}

\label{fig:resol}
\end{figure*}
%%%%%%%%%%%%%%%%%%%%%%%%%%%%%%%%%%%%%%%%%%%%%%%%%%%%%%%%%%%%%%%

\subsection{X-ray mirror \phantom{\huge I}}

The grazing-incidence mirror will consist of many pairs (for two
reflections) of thin monocrystalline silicon shells, coated with either Ir
or Pt to provide X-ray reflectivity. The mirror's 1.5m diameter and 4m focal
length are selected to provide a large effective area at the energies of
interest ($E<2$ keV) and a broad angular field of view, covering a Milky Way
like galaxy at $z=0.01$ (40 Mpc), to fit within the detector of a feasible
linear size. This choice of geometry means that the effective area at higher
energies is low. The mirror is being developed at NASA GSFC. The angular
resolution requirements imposed by the main science drivers is modest,
10\as\ half-power diameter (HPD) or 15\as\ square pixel enclosing 90\%
energy. A resolution of 2.8\as\ HPD has already been demonstrated in the lab
for a small subset of the mirror (Fig.\ \ref{fig:psf}). The mirror is
designed to have a constant PSF and insignificant vignetting across the
30\am\ field of view.

\subsection{Detector}

The \LEM\ detector is a cryocooled array of Transition-Edge Sensor (TES)
microcalorimeters. It will be built at NASA GSFC using the technology
developed for \athena\ XIFU and the \lynx\ concept by the same
team\cite{Lee15,Miniussi18}. The array will consist of 13,806 absorber
pixels (equivalent of $118\times 118$ square) with a 290\,$\mu$m pitch, in
hexagonal arrangement. For a 4m focal length, it will cover a solid angle
equal to a 29.4\am\ square.

The inner 1062 absorber pixels (a $7'\times 7'$ square array) will each be
connected to a single TES and provide an energy resolution of 1 eV
(FWHM). The rest of the array will consist of ``hydras,'' where $2\times2$
absorber pixels are connected to a single TES, which will determine which
absorber was hit with a photon based on the different time constants for the
different absorbers\cite{Smith20}. These pixels will provide a resolution of
2 eV (FWHM) for $E<2$ keV. The absorber thickness will be optimized for the
low X-ray energies and the sensor will take advantage of the narrow energy
band of the mirror. Figure \ref{fig:resol} shows the measured line response
for a prototype \lem\ array\cite{Smith23}, which already nearly meets our
requirements both for the single pixels and the hydras. The position
discrimination within the hydra has also been demonstrated at least for
$E>300$ eV.\cite{Smith23}

\lem\ will use the IR and optical blocking filters developed for
\athena\ XIFU, scaled up in size for a wider viewing cone of the
\lem\ mirror.

The calorimeter array is cooled to the $35-50$\,mK operating temperature by the
Adiabatic Demagnetization Refrigerators built at NASA GSFC and the
cryocooler built by Lockheed Martin.

A modulated X-ray source will operate in orbit and illuminate the array with
a 1.5 keV Al K$\alpha$ near-monochromatic line for continuous calibration of
the energy response. The resulting accuracy on the pixel energy gain, 0.25
eV at 1.5 keV (and proportionally lower at lower energies), corresponds to a
50 km/s accuracy (1$\sigma$) on the relative velocity measurements using the
Doppler shift.

\vspace*{1mm}
\subsection{Spacecraft and orbit}

The spacecraft will be built by Lockheed Martin Corp. The main mode of
operations will be long pointed observations. \lem\ will also be able to
slew to perform efficient scans of large areas of the sky --- such as the
whole sky (\S\ref{sec:lass}). While \lem\ is not a time-domain mission and
not required to have a fast response capability, it will be able to respond
to the highest-profile events on a several-day timescale from first notice
to pointing --- similar to \chandra. The area of the sky accessible at any
particular moment will be approximately half the sky; this fraction is
currently under study.

The orbit for \lem\ is currently under study. The Sun-Earth Lagrange points
L1 and L2 would provide a stable thermal environment for the
microcalorimeter and the high observing efficiency to take full advantage of
the 5 yr mission minimum lifetime. A possibility to extend the field of
regard to include observations of the Earth's magnetosphere from the L1
orbit is being considered.

\vspace*{1.5mm}
\section{SCIENCE DRIVERS}
\label{sec:sci}

Below is a brief description of the science investigations that will be
enabled by \lem. They will be presented in more detail in several
forthcoming papers.

\subsection{Physics of galaxy formation}
\label{sec:galform}

Galaxies form from the collapse of primordial matter density fluctuations
under the effect of gravity, which is counteracted by feedback from stars
and the central supermassive black holes that pushes the gas out of the
galaxy. The result is a circulation of gas and metals (synthesized by stars
and supernovae in galaxies) into and out of galaxies --- a galactic
ecosystem. Understanding this ecosystem is key to understanding galaxy
formation. Quoting Astro2020, ``the key missing link in unveiling the
physics driving galaxy growth is to measure the properties of the diffuse
gas within, surrounding, and between galaxies,'' and these three locations
constitute the main \lem\ science drivers.

%%%%%%%%%%%%%%%%%%%%%%%%%%%%%%%%%%%%%%%%%%%%%%%%%%%%%%%%%%%%%%%
\begin{figure}[b]
\vspace*{1mm}
\centerline{%
\includegraphics[width=0.48\textwidth]{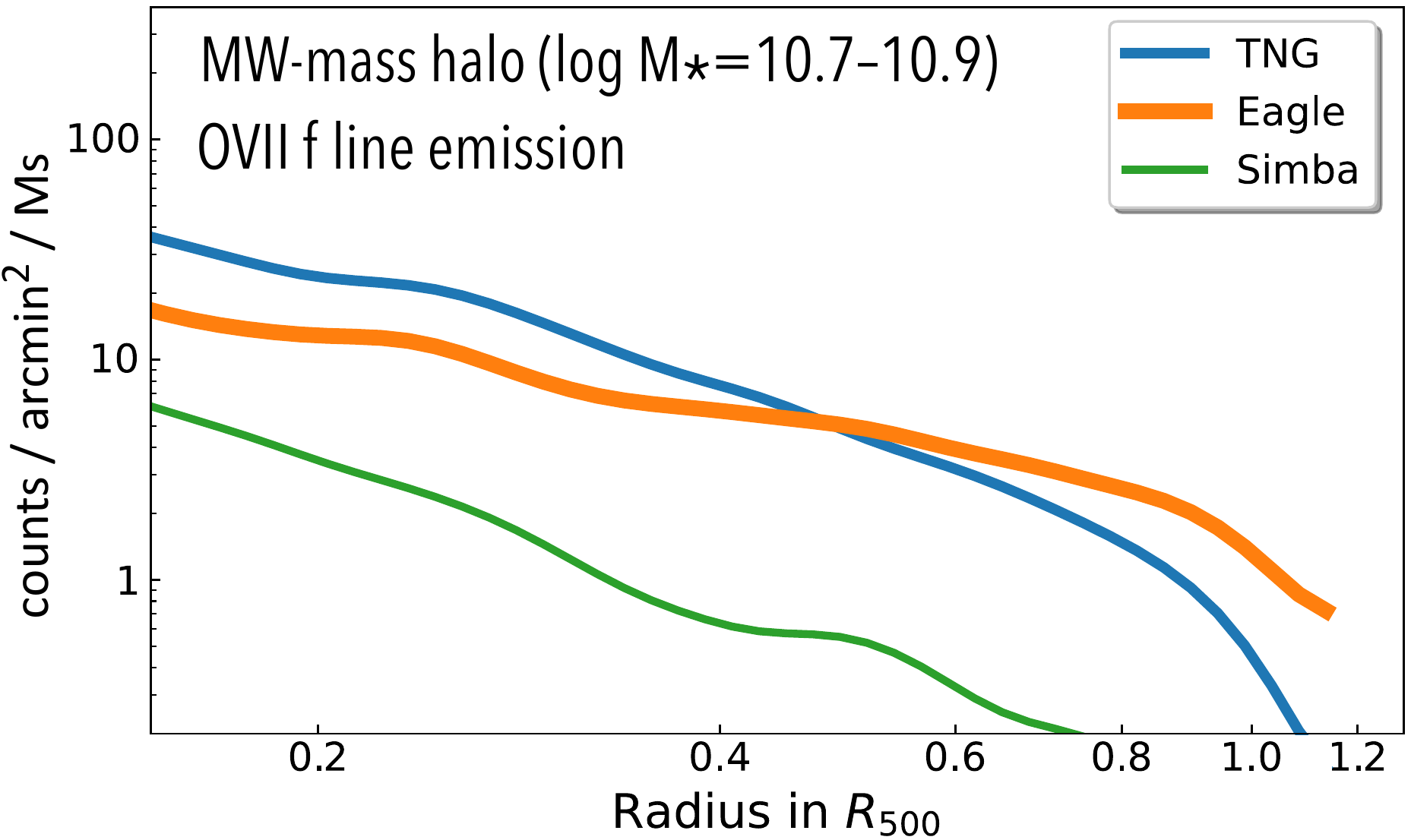}}

\vspace{1mm}
\caption{Predicted brightness profiles of the O{\sc vii} emission for Milky
Way mass galaxies from three cosmological simulations (TNG\cite{Nelson19a},
EAGLE\cite{Schaye15,Crain15}, Simba\cite{Dave19}).  O{\sc vii} is abundant
at $T=(0.3-2)\times 10^6$\,K that is characteristic of CGM. Each curve is a
median profile for 50 simulated galaxies. The expected \lem\ counts are
shown for a 2 eV wide interval centered on the forbidden line of the O{\sc
vii} triplet at $z=0.01$.  $R_{200}\simeq1.5R_{500}$. These numerical models
include different physical prescriptions for feedback, and the resulting
predicted X-ray brightness at large radii differs by over an order of
magnitude. The CGM extent and spatial structure are sensitive diagnostics of
the feedbak physics; detecting and mapping this emission will strongly
constrain the physical processes shaping the galaxies.}

\vspace*{-2mm}
\label{fig:profs}
\end{figure}
%%%%%%%%%%%%%%%%%%%%%%%%%%%%%%%%%%%%%%%%%%%%%%%%%%%%%%%%%%%%%%%

%%%%%%%%%%%%%%%%%%%%%%%%%%%%%%%%%%%%%%%%%%%%%%%%%%%%%%%%%%%%%%%
\begin{figure*}[tb]
\centerline{%
\includegraphics[width=0.95\textwidth]{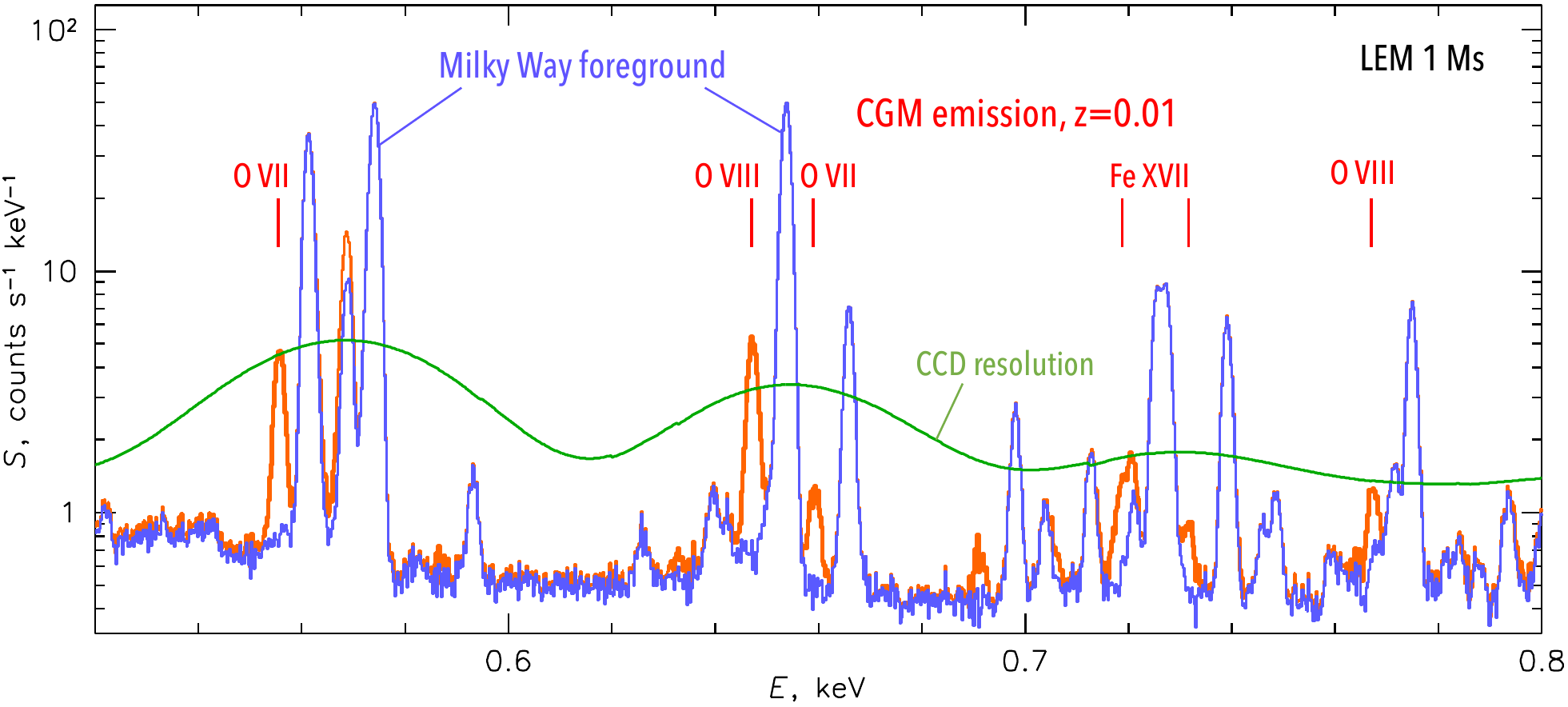}}

%\vspace{5mm}
\caption{Simulated \lem\ spectrum of the CGM of a galaxy twice as massive as
the Milky Way (from TNG simulations\cite{Nelson19a}), at $z=0.01$ or 40 Mpc, for
a 1 Ms exposure and the 2 eV resolution of the full detector.  Blue line
shows the spectrum of the foreground, red line includes the CGM signal. The
spectrum is extracted from a broad annulus between
$0.25-0.75\,R_{500}$. Point sources detected with the \lem\ angular
resolution are excluded. Because the Milky Way emits in the same spectral
lines but is over an order of magnitude brighter, it is impossible to
disentangle the two with the present CCD instruments such as \Chandra\ and
\XMM\ (green line). A nondispersive spectrometer with 2 eV resolution allows
the study of the faint CGM for galaxies at $z\gax 0.01$, where they can also
be well-resolved spatially.}

\label{fig:cgmspec}
\end{figure*}
%%%%%%%%%%%%%%%%%%%%%%%%%%%%%%%%%%%%%%%%%%%%%%%%%%%%%%%%%%%%%%%

%%%%%%%%%%%%%%%%%%%%%%%%%%%%%%%%%%%%%%%%%%%%%%%%%%%%%%%%%%%%%%%
\begin{figure*}[bt]
%\vspace*{3mm}
\centerline{%
\includegraphics[width=0.99\textwidth]{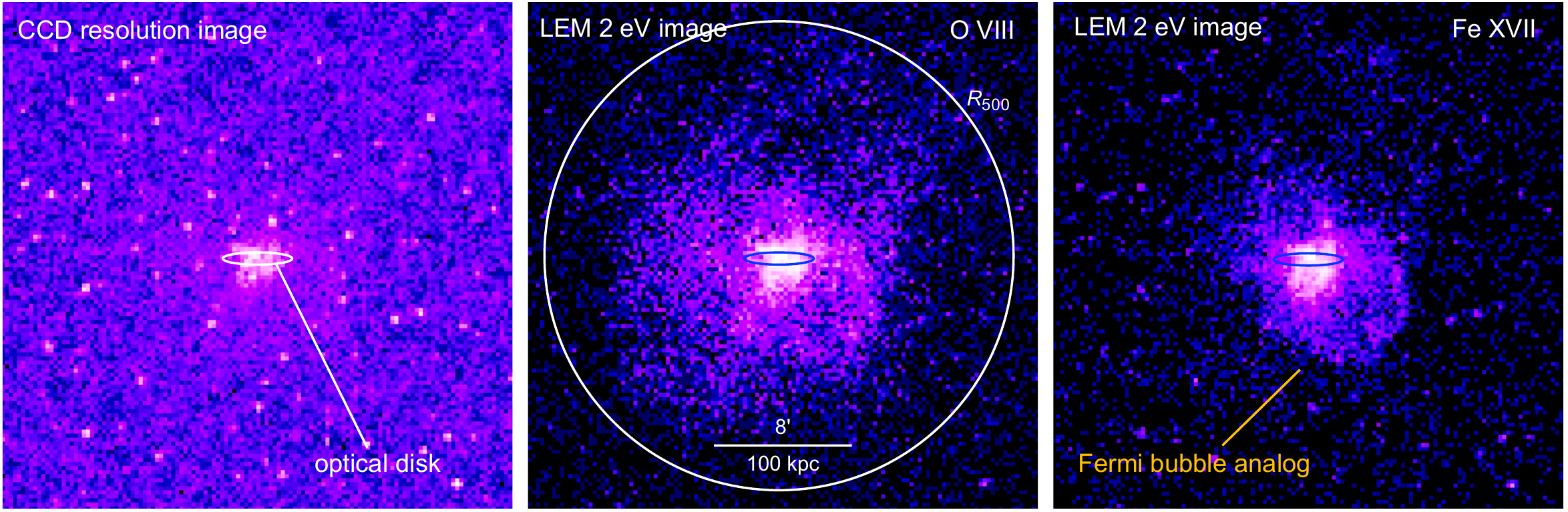}}

\vspace{4mm}
\caption{Simulated Milky Way-like galaxy at $z=0.01$
(TNG50\cite{Pillepich19,Pillepich21,Nelson19b} of the IllustrisTNG
simulation suite). Panels are 30\am\ with 15\as\ pixels (\LEM\ field of view
and angular resolution). The ellipse shows the size of the optical disk,
seen edge-on. {\em Left}: Soft X-ray image with a CCD-like (70 eV) spectral
resolution centered at the O{\sc viii} line, including the Milky Way
foreground and CXB (the brightest sources accounting for half of CXB flux
have been spatially removed). The bright foreground makes it impossible to
detect extended CGM emission. {\em Middle, Right}: \LEM\ 1 Ms images of the
same galaxy, but in 2 eV wide bins (the \lem\ spectral resolution) centered
on the O{\sc viii} and Fe{\sc xvii} CGM emission lines. The MW foreground is
almost completely resolved away, taking advantage of the galaxy's known
redshift; the use of the narrow bin also removes most of the residual
power-law CXB flux. \LEM\ unveils the rich spatial structure of the
line-emitting CGM halo --- including the analog of the Fermi/eROSITA bubbles
produced by the central SMBH. Mapping different emission lines provides a
measurement of the plasma temperature.}

\label{fig:3halos}
\end{figure*}
%%%%%%%%%%%%%%%%%%%%%%%%%%%%%%%%%%%%%%%%%%%%%%%%%%%%%%%%%%%%%%%

%%%%%%%%%%%%%%%%%%%%%%%%%%%%%%%%%%%%%%%%%%%%%%%%%%%%%%%%%%%%%%%
\begin{figure*}[tp]
\centerline{%
\includegraphics[width=0.99\textwidth]{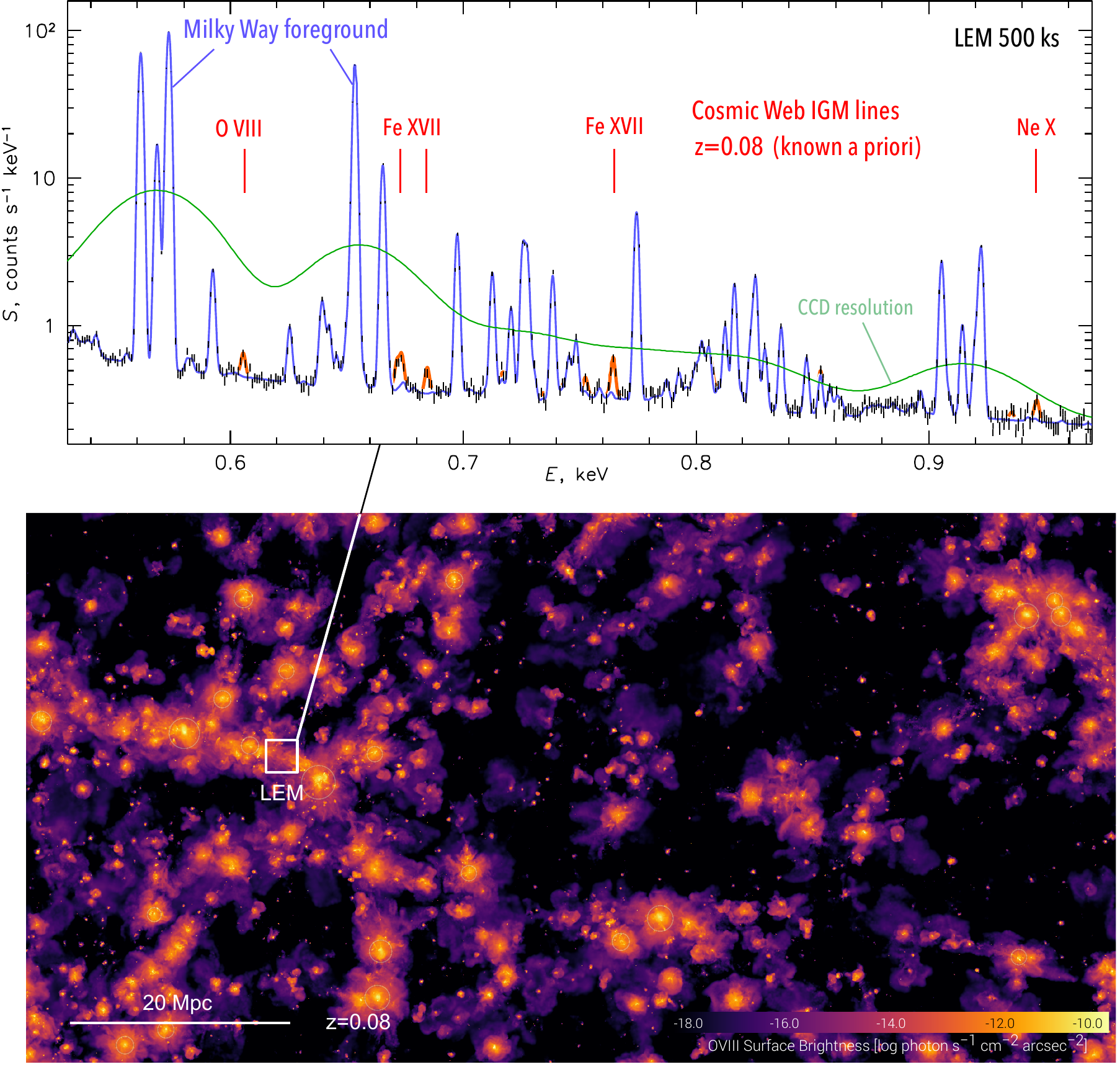}
}

\vspace{5mm}
\caption{Simulated \LEM\ spectrum of the Intergalactic Medium (IGM) in the
Cosmic Web. Owing to its high spectral resolution and large grasp,
\LEM\ will be able to detect emission from metals in the denser regions of
the Cosmic Web filaments, outside the confines of galaxies (and their CGM
halos) and galaxy clusters. It is a diagnostic of galaxy formation processes
at early epochs. Filaments will be selected from galaxy surveys and the
redshift of the IGM will be known, i.e., it is not a blind search for
emission lines.  The image shows O{\sc viii} emission in cosmological
simulations (TNG100\cite{Nelson18}). The sample spectrum corresponds to a
typical bright region of an IGM filament near a massive cluster (at a
distance $r\sim 2R_{200}$ from the cluster), excluding halos of all galaxies
and clusters. The eV spectral resolution is essential for separating the
faint IGM lines (highlighted by red) from the much brighter Milky Way
foreground (cf.\ the green line that shows the same spectrum at CCD
resolution).}

\vspace{10mm}
\label{fig:igm}
\end{figure*}
%%%%%%%%%%%%%%%%%%%%%%%%%%%%%%%%%%%%%%%%%%%%%%%%%%%%%%%%%%%%%%%

{\bs CGM.}~~At the mass scale of the Milky Way, the observable galaxy
properties exhibit a dramatic change --- from star-forming disks thought to
be fed by cold gas streams, to passive ellipticals surrounded by hot
hydrostatic gas. Apparently, this reflects the shift in the balance of the
various galaxy-shaping physical processes. Galaxies around this mass should
thus offer a particularly sensitive probe of those processes. Indeed,
cosmological simulations of the MW-mass galaxies that include different
physics make strikingly different predictions for the X-ray properties of
the outer regions of the circumgalactic halos (Fig.\ \ref{fig:profs}). At
present, these predictions are entirely unconstrained by observations, and
detecting and measuring this X-ray signal can tell us which physical
processes shape the CGM.

\lem\ will be able to map the CGM (defined here as the gaseous halo within
the galaxy virial radius $R_{200}$) in the emission lines of the dominant
ion species, O{\sc vii}, O{\sc viii} and Fe{\sc xvii}. The high spectral
resolution will allow us to disentangle the faint signal from the CGM of
external galaxies from the overpowering Milky Way foreground using the
galaxy's redshift. Because the Milky Way itself has a CGM halo that emits in
the same atomic transitions --- and we are immersed in its bright central
region and see that emission across the sky --- the ability to separate the
redshifted extragalactic signal by energy is a critical advantage of \lem,
as illustrated in Fig.\ \ref{fig:cgmspec}. The current X-ray CCD detectors
lack spectral resolution for this approach.

For the Milky Way mass halos, \lem\ will be able to map the CGM emission
{\em for individual galaxies}\/ (as opposed to stacking the radial profiles
of many galaxies) in the brightest emission lines out to $R_{500}$ and
beyond, and derive the distribution of gas temperature, density, pressure
and elemental abundances out to a large fraction of $R_{500}$. The maps are
expected to reveal a wealth of spatial structure, possibly including the
extragalactic analogs of our own Milky Way's Fermi bubbles
(Fig.\ \ref{fig:3halos}) --- if black hole / AGN feedback is as important as
theory predicts.  Such measurements are only possible because the Milky Way
foreground can be separated using the redshift of the signal. The minimum
redshift for such separation for the 2 eV detector resolution is $\sim 0.01$
(Fig.\ \ref{fig:cgmspec}); at this redshift, the $R_{500}$ for a Milky Way
mass galaxy fills the \lem\ FOV and the whole galaxy can be mapped in one
deep pointing. The \lem\ angular resolution will allow efficient spatial
masking of the background point sources and resolving the interesting linear
scales in the halo (e.g., shocks around the Fermi bubbles). A typical CGM
halo is optically thin for the emission lines of interest over most of its
volume except for the very central regions, where the optical depth for the
O{\sc vii} resonant line becomes significant. \lem\ will easily resolve the
components of the O{\sc vii} He$\alpha$ triplet (where the resonant line is
affected by scattering while other components are not), which will enable
interesting physical diagnostics of the gas.

Using the high-resolution (1 eV) central array, \LEM\ will be able to
simultaneously map the velocities in the brighter central regions of the
galactic halos. It will detect gas bulk velocities and velocity dispersion
predicted for the particularly energetic modes of feedback.

%%%%%%%%%%%%%%%%%%%%%%%%%%%%%%%%%%%%%%%%%%%%%%%%%%%%%%%%%%%%%%%
\begin{figure*}[tb]
\centerline{%
\includegraphics[width=1.0\textwidth]{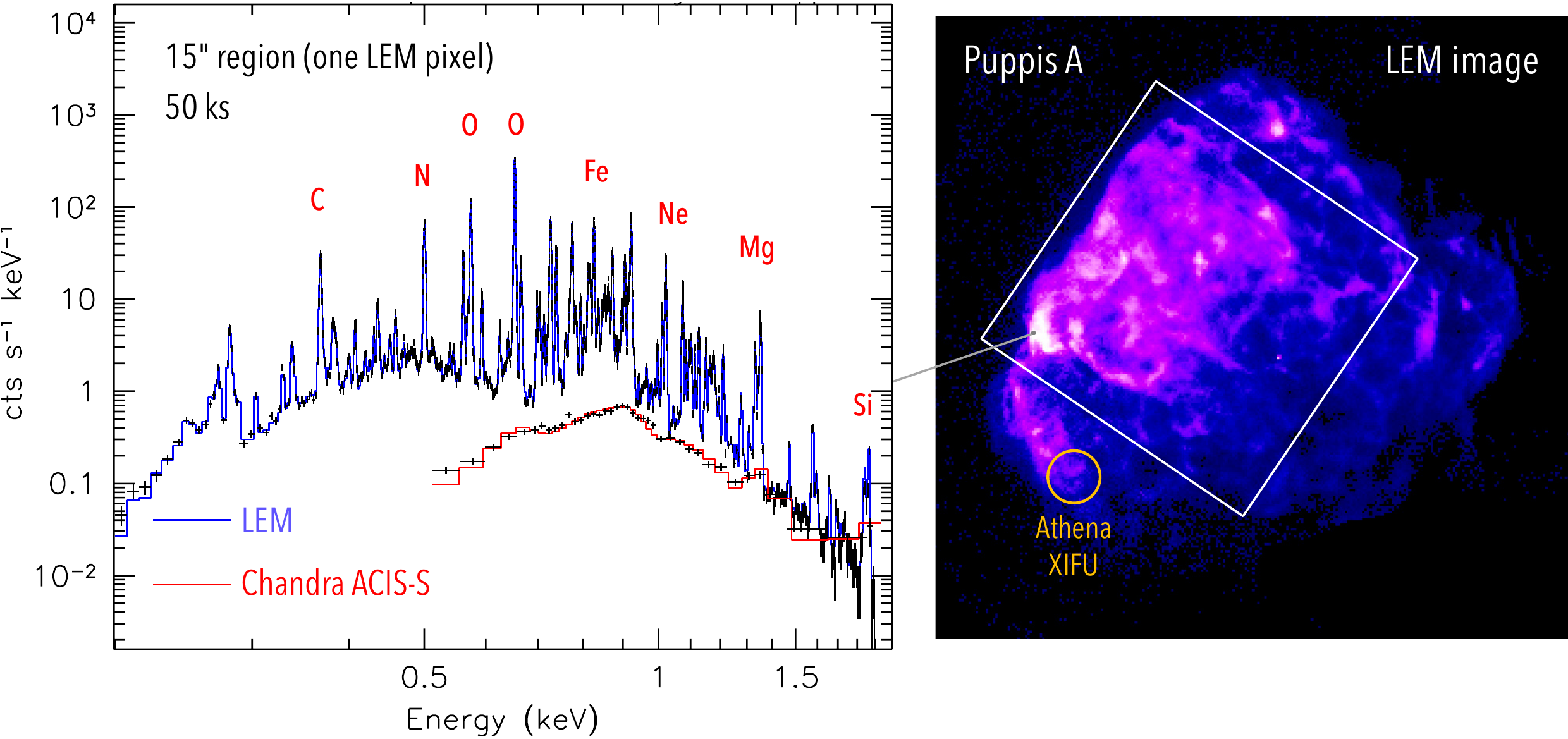}
}

\vspace*{-1mm}
\caption{{\em Left}: Sample spectrum from one pixel in the \LEM\ calorimeter
from a 50 ks exposure of the Puppis A supernova remnant. {\em Right}:
\LEM\ image of Puppis A. The white square shows one \LEM\ FOV and the circle
shows the \Athena\ XIFU FOV. \LEM\ will map such large SNRs very
efficiently, which is especially valuable for old, large Galactic SNRs. Each
\lem\ pixel will provide an extremely line-rich spectrum, resolving multiple
lines for each ion, allowing us to map chemical composition, diagnose
nonequilibrium conditions in the plasma and measure velocities of ejecta. A
CCD spectrum (\chandra\ ACIS) is shown for comparison; all spectral features
are blended. A dispersive spectrometer (gratings) provides very limited
spectral resolution for such extended sources because of spatial blending.}

\label{fig:pupa}
\end{figure*}
%%%%%%%%%%%%%%%%%%%%%%%%%%%%%%%%%%%%%%%%%%%%%%%%%%%%%%%%%%%%%%%

{\bs IGM.}~~An even more challenging observing target is the IGM, which we
define as the diffuse matter outside $R_{200}$ of any galactic or cluster
halos. The warm-hot phase of it ($>10^5$\,K) is known as WHIM. It is the
ultimate repository of everything expelled from the galaxies over their
lifetime, and its metal content thus encodes the history of the feedback
processes in the Universe. The IGM regions immediately surrounding massive
hot galaxy clusters, where the gas temperature is still above $10^7$\, are
accessible for sensitive low-background CCD instruments, which could use the
emission at $E>1$ keV.  However, most of the IGM has lower temperatures, and
its emission will be flooded by the bright Milky Way. \lem\ will be able to
detect the brightest emission lines from the dense regions of the Cosmic Web
filaments that connect galaxy clusters, at distances $2R_{200}$ from the
clusters and beyond.

We will use optical galaxy redshift surveys to select such filaments at an
optimal redshift ($z=0.07-0.08$) to be able to see the most promising IGM
emission lines through the Milky Way line forest (Fig.\ \ref{fig:igm}) and
survey these regions with multiple deep pointings. This is where the large
grasp is critically important, as the number of photons collected from a
very extended source is directly proportional to it --- and the IGM is
extremely faint. While we will not be able to detect the {\em continuum}\/
emission from the IGM (it is too far below the background), we will detect
the faint emission lines of O{\sc vii}, O{\sc viii} and Fe{\sc xvii} (if,
that is, the IGM is indeed enriched, as the current theory predicts), and
compare the brightness in these lines as a function of the galaxy number
density with cosmological simulations, in order to determine when the
enrichment of the IGM by galactic ejecta has occurred and what physical
processes were responsible.

\lem's large grasp and good angular resolution will allow us to detect the
IGM {\em in absorption} in the cumulative spectra of the CXB sources in the
same fields where we will search for the IGM emission. A comparison of
emission and absorption for the same filament will allow us to derive the
physical properties of the filament gas.

The line emission from the low-density IGM can be significantly enhanced (in
the resonant transitions only), compared to intrinsic thermal emission, by
resonant scattering of the CXB photons, if the CXB point sources are excised
from the image\cite{Khabibullin19}. This improves the
detectability of the IGM emission. This also applies to the outer regions of
the CGM, with a significant additional enhancement from the resonant
scattering of the photons from the galaxy's own bright central region. This
will extend the radius at which the CGM could be detected by \lem. Of
course, interpretation of the detected signal will require modeling of the
scattered contribution; \lem\ will be well equipped for such spectral
diagnostics because it will resolve the resonant and forbidden transitions
for the relevant ions.

{\bs Abundances in the ICM.}~~ \lem\ will address the cumulative effect of
the galactic feedback from another angle. The ratios of the various elements
in the ICM in the outskirts of galaxy clusters, as well as the total metal
content of the ICM and the spatial distribution of metals in the ICM, depend
on the prevalent star populations and types of supernovae that generated
those metals, when it happened, and how the metals mixed into the
intergalactic gas\cite{Mernier22}. \lem\ will map the
abundances and abundance ratios for O and Fe, but also for a number of
less-abundant elements that can be used as sensitive diagnostics for the
sources of chemical enrichment. \lem's large grasp and good angular
resolution will allow us to efficiently and completely map the abundances of
all the interesting elements in nearby groups and clusters in the
interesting radial range $0.5R_{500}-R_{200}$.

{\bs Velocities of the ICM.}~~ \lem\ will efficiently and completely map the
gas velocities and velocity dispersions in nearby clusters and groups,
relaxed and dynamic, out to $R_{200}$ (using mostly the Fe-L lines). It will
determine where the clusters are hydrostatic and where the contribution of
turbulence and gas bulk motions into the total energy budget becomes
important. This information is key for our understanding of the physics of
the large scale structure and cluster formation.

%%%%%%%%%%%%%%%%%%%%%%%%%%%%%%%%%%%%%%%%%%%%%%%%%%%%%%%%%%%%%%%
\begin{figure*}[tp]
\centerline{%
\includegraphics[width=1.0\textwidth]{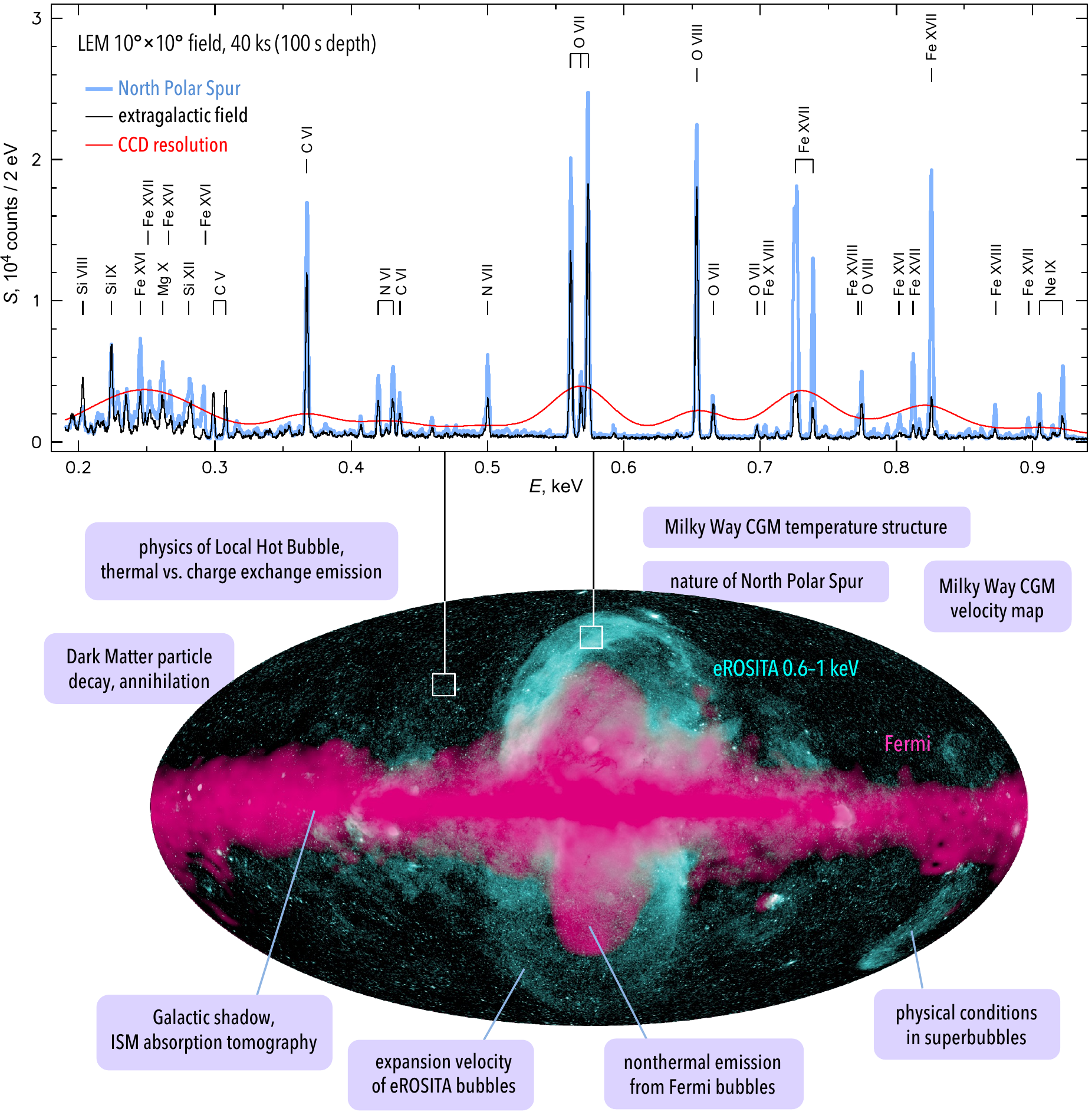}
}

\vspace{5mm}
\caption{\LEM\ shallow all-sky survey. The all-sky map\cite{Predehl20} in
Galactic coordinates overlays soft X-ray emission from {\em SRG}/eROSITA
(cyan) and $\gamma$-ray emission from \fermi\ (magenta). {\em Top}: Example
simulated spectra for $10^\circ\times 10^\circ$ regions (marked by squares)
and 100s depth are shown for an average high-latitude region outside of any
bright features (``extragalactic field''), and another for the North Polar
Spur; the CXB is included. Purely thermal CIE plasma models are assumed. Red
curve shows the NPS spectrum at CCD resolution (e.g., eROSITA).  \lem\ will
resolve the forest of lines and gain access to line diagnostics for
temperature, nonequilibrium and charge exchange processes.
The survey will allow us to map the temperature structure and velocities of
the Milky Way inner CGM on few-degree scales --- including the expansion of
the eROSITA/Fermi bubbles, believed to be produced by the central SMBH or
star-forming regions. Numerous other investigations, some of which are
labeled in the figure, will become possible.}

\label{fig:allsky}
\end{figure*}
%%%%%%%%%%%%%%%%%%%%%%%%%%%%%%%%%%%%%%%%%%%%%%%%%%%%%%%%%%%%%%%

{\bs Stellar feedback}. \lem\ will study the gas inside the Milky Way that
is part of the galactic ecosystem. It will map the chemical composition and
velocities of ejecta in well-resolved supernova remnants
Fig.\ \ref{fig:pupa}). It will also map the physical conditions in the Milky
Way and Local Group superbubbles --- regions where stellar winds from many
massive stars and the supernovae combine and form large cavities in the ISM
filled with hot plasma. If superbubbles break, the hot plasma can escape
from the galactic disk, forming galactic outflows or fountains and flowing
into the CGM. This is the mechanism by which the stellar feedback in
galaxies operates. Mapping the X-ray line emission will offer diagnostics of
gas out of collisional ionization equilibrium (rapidly heating/cooling) and
separating thermal emission from the charge exchange emission that arises
from the mixing of cold and hot gas phases. With its large grasp, \lem\ will
be able to efficiently map very extended structures as Orion-Eridanus and
other nearby superbubbles.

\subsection{All-Sky Survey \phantom{\huge I}}
\label{sec:lass}

As described above, the IGM and the CGM of external galaxies are observed
through the forest of bright Milky Way emission lines. It will be helpful to
find areas in the sky where this foreground is less bright, at least in
certain problematic lines such as Fe{\sc xvii} and Fe{\sc xviii}. (This is
similar to \Hubble\ peering into the Lockman Hole to observe high-$z$\/
galaxies.) The \erosita\ and \halosat\ sky surveys will be helpful, but lack
spectral resolution to detect faint emission lines. \lem\ has sufficient
grasp to perform its own shallow all-sky survey in a relatively modest total
exposure --- e.g., an all-sky coverage with a useful 10s depth will require
1.6 Ms. We envision performing such shallow pathfinder survey very early in
the mission to identify areas in the sky best suited for deep CGM/IGM
exposures and for other interesting studies, and add up to 10 times more
depth over the mission lifetime.

In addition to the above technical purpose, the first-ever all-sky survey
with an imaging calorimeter will open {\bs enormous discovery space}. With
\lem's large effective area, even a shallow depth will be sufficient for
groundbreaking studies.  The statistical quality and richness of the spectra
for typical low-brightness and high-brightness regions of the sky for the
eventual full depth are illustrated in Fig.\ \ref{fig:allsky}. The figure
also lists examples of investigations using all-sky or partial-sky data that
we can envision at present.

Some important aspects of the {\bs feedback processes} in Milky-Way mass
galaxies would be best viewed from inside the Milky Way. \lem\ will map the
velocities of the inner regions of the Milky Way CGM, and in particular, the
expansion of the Fermi/eROSITA bubbles, believed to be evidence of feedback
from either the SMBH or star-forming regions in the Galactic Center. It will
map the temperature structure of the inner CGM across the sky and along the
line of sight using lines of the various ion species --- something only a
calorimeter can do, in the presence of multiple temperature components and
solar wind charge exchange on each line of sight. The MW observations will
complement the studies of CGM in other galaxies (\S\ref{sec:galform}), where
\lem\ will map the outer halos but have limited insight into the interface
between the disk and the halo, where the exchange of mass and metals with
the outer CGM takes place.

The emission of the lower MW halo (thought to be formed from the exhaust of
Galactic chimneys) can be separated from the emission of the Galactic disk
ISM by observing a dense absorbing extraplanar cloud, and determining the
spectra of the foreground and background to the cloud. \LEM's large FOV and
high spectral resolution will allow the first definitive separation of the
disk and halo spectra for individual plasma diagnostics.

\lem's sensitivity and spectral resolution at low energies will let us
unambiguously separate the the heliospheric solar wind charge exchange
(SWCX) line emission from thermal emission of the Local Hot Bubble (LHB) and
the Galactic halo. This will let us explore the physics of the LHB. The
North Polar Spur is another enigmatic structure in the Milky Way sky,
possibly projected onto the northern Fermi bubble (Fig.\ \ref{fig:allsky});
its spectrum is known (at CCD resolution) to be very different from the rest
of the sky above the Galactic plane. A survey with \lem\ will provide the
plasma temperature and velocity map for this feature, as well as any
possible contribution from the charge exchange mechanism, and may explain
its nature.

The Fermi bubbles (as distinct from the eROSITA bubbles that surround them,
Fig.\ \ref{fig:allsky}) contain cosmic ray electrons that generate the
synchrotron ``microwave haze'' observed by {\em WMAP}\/ and {\em Planck}\/
and the inverse Compton $\gamma$-ray emission discovered by
\fermi\cite{Ackermann14}. The synchrotron radiation generated by the
highest-energy end of this population may be detectable in the soft X-ray
band covered by \lem. \lem\ is a very sensitive instrument for detecting
faint {\em continuum}\/ emission at $E<1$ keV, because it can resolve away
the emission lines that dominate the soft X-ray sky. \lem\ will have the
requisite grasp to map the entire Fermi bubbles as part of the all-sky
survey and yield constraints on the X-ray synchrotron emission from the
southern Fermi bubble. \LEM\ upper limits --- or a positive detection ---
will provide insights into the energetics of the Fermi bubbles.

%%%%%%%%%%%%%%%%%%%%%%%%%%%%%%%%%%%%%%%%%%%%%%%%%%%%%%%%%%%%%%%
\begin{figure*}[t]
\centerline{%
\includegraphics[width=0.99\textwidth]{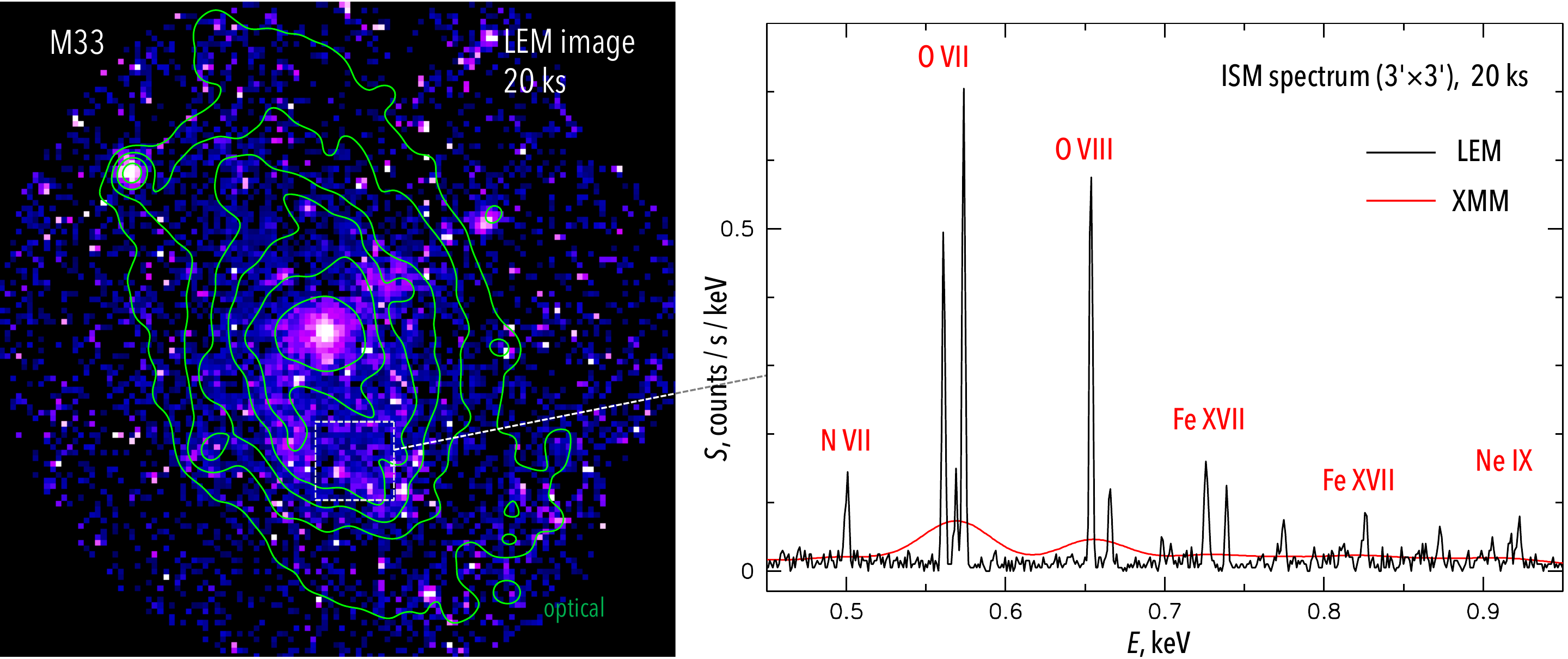}
}

\caption{\LEM\ will be able to spatially separate and perform spectroscopy
of the ISM and the X-ray binaries in nearby spiral galaxies, such as M33. A
simulated 20 ks \lem\ image (based on the existing \xmm\ observation) and a
sample ISM spectrum are shown. \lem\ will detect many binaries and probe
their variability. It will also produce detailed maps of the line-dominated
ISM emission. For such nearby galaxies, the Milky Way foreground cannot be
separated using the redshift, but its contamination for most ISM lines is
under 10\%. In each observation, \lem\ will provide the same amount of
imaging information as \xmm, but 20--30 times more spectral information.}

\label{fig:m33}
\end{figure*}
%%%%%%%%%%%%%%%%%%%%%%%%%%%%%%%%%%%%%%%%%%%%%%%%%%%%%%%%%%%%%%%

\subsection{Observatory science \phantom{\Large I}}
\label{sec:obs}

In addition to the main science drivers described above (which will be
addressed during the 30\% of the mission observing time allocated to the
science team), \lem\ will transform all areas of X-ray astrophysics. It is
useful to think of \lem\ as an \xmm\ with a calorimeter --- its imaging
capabilities (FOV, angular resolution, effective area) are similar in the
$E<2$ keV band to EPIC's, while its spectral resolution is 50 times
better. Many studies will be serendipitous to the long CGM and IGM
exposures, while others will be proposed by general observers (GO) for the
remaining 70\% of the observing time. Many of them will need only modest
exposures. An incomplete sample of possible novel short investigations is
given below.

\smsk {\bi Cooling, AGN feedback, gas mixing}\/ in cool cores of galaxy
clusters. Where is the coolest gas in the Perseus cluster core and how cool
does it get? \lem\ will use its 1 eV resolution in the central array to
derive the temperature distribution for the cooling gas. It will also map
the charge exchange emission that may come from the interface between the
hot ICM and the molecular gas nebula. If such emission is detected and
spatially correlated with the cold nebula (for which at least a \lem-like
angular resolution is required), it would provide invaluable insight into
the process of gas mixing --- a critical piece of the puzzle of the energy
balance in the cluster cool cores.

\smsk {\bi Power spectrum of turbulence}\/ in cool galaxy clusters and
groups using the velocity structure function. Because of its large grasp,
\lem\ will be able to cover the entire cluster in a short exposure. The
full spatial coverage greatly increases the accuracy of the structure
function and the power spectrum measurement.\cite{ZuHone16} A power spectrum
of turbulence will provide the measure of the energy flux from the cluster
mergers down the turbulent cascade to the dissipation scale.

\smsk {\bi Protoclusters at high z.}\/ Deep exposures toward IGM and CGM
fields will return significant numbers of serendipitous distant galaxy
clusters, for which \lem\ will be able to determine the redshift using its
own X-ray spectra. Dedicated surveys can be also devised. \lem\ will derive
chemical abundances for high-$z$\/ clusters and constrain the cosmological
enrichment history.

\smsk {\bi High-z AGN.} Every \lem\ deep field will have multiple AGN at
$z>2.5$ with a detectable Fe K$\alpha$ line. \LEM\ will measure the profiles
of these lines to study broad-line region clouds and accretion disk
reflection. It will study heavily obscured AGN and expand significantly upon
the AGN science derived from current \chandra\ and \xmm\ deep fields.

\smsk {\bi AGN outflows and feedback.} With its superior energy resolution,
\lem\ will be able to precisely map fluorescent and photoionization lines
and decouple AGN signatures in emission and absorption from ionization by
stellar sources to understand the role that AGN outflows and feedback play
in galaxy evolution.

%%%%%%%%%%%%%%%%%%%%%%%%%%%%%%%%%%%%%%%%%%%%%%%%%%%%%%%%%%%%%%%
\begin{figure*}[t]
\centerline{%
\includegraphics[width=0.9\textwidth]{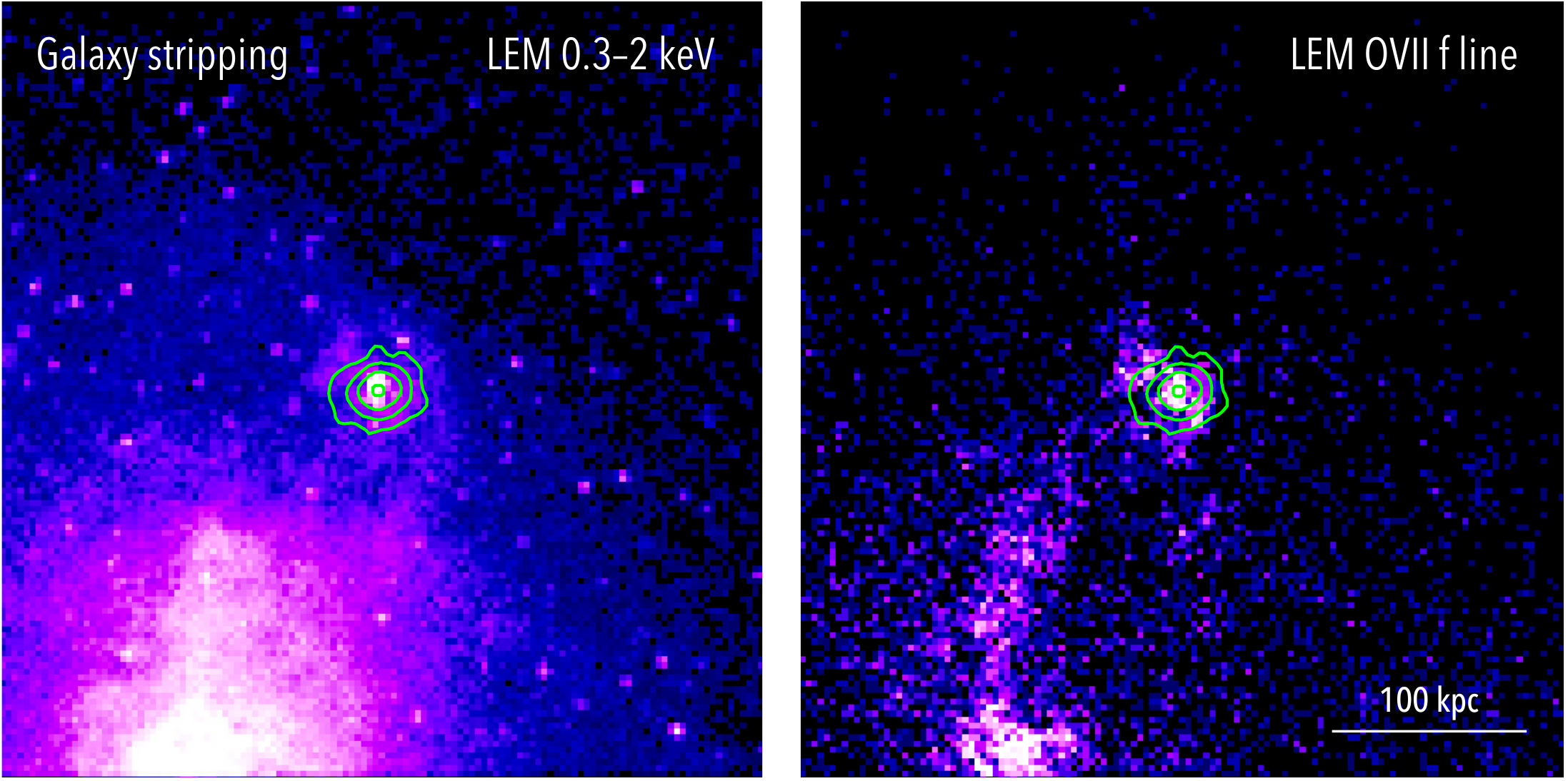}
}

\caption{\LEM\ will map the process of galaxy stripping and mixing of the
stripped gas in detail.  A 100 ks \lem\ exposure of interacting galaxies at
$z=0.01$ (TNG100 simulations\cite{Yun19}) --- an $M_{\rm tot}=2\times
10^{11}$\msun\ galaxy (at the panel center, with contours showing optical
light) after the passage through a group-mass halo (at the lower edge of the
panels). The CXB is included, except for its brightest point sources, which
are removed. {\em Right}: \lem\ image in a narrow 3 eV interval at the
redshifted O{\sc vii} line (the forbidden component of the triplet, which at
this redshift can be resolved from the corresponding Milky Way foreground
line as seen in Fig.\ \ref{fig:cgmspec}) reveals a ``jellyfish'' --- a trail
of cool gas ram-pressure stripped from the infalling galaxy. The hotter
group halo does not emit in this line. Such gas tails are a sensitive
diagnostic of the plasma physics in galaxy groups and clusters.}

\label{fig:jellyfish}
\end{figure*}
%%%%%%%%%%%%%%%%%%%%%%%%%%%%%%%%%%%%%%%%%%%%%%%%%%%%%%%%%%%%%%%

%%%%%%%%%%%%%%%%%%%%%%%%%%%%%%%%%%%%%%%%%%%%%%%%%%%%%%%%%%%%%%%
\begin{figure*}[t]
\centerline{%
\includegraphics[width=0.97\textwidth]{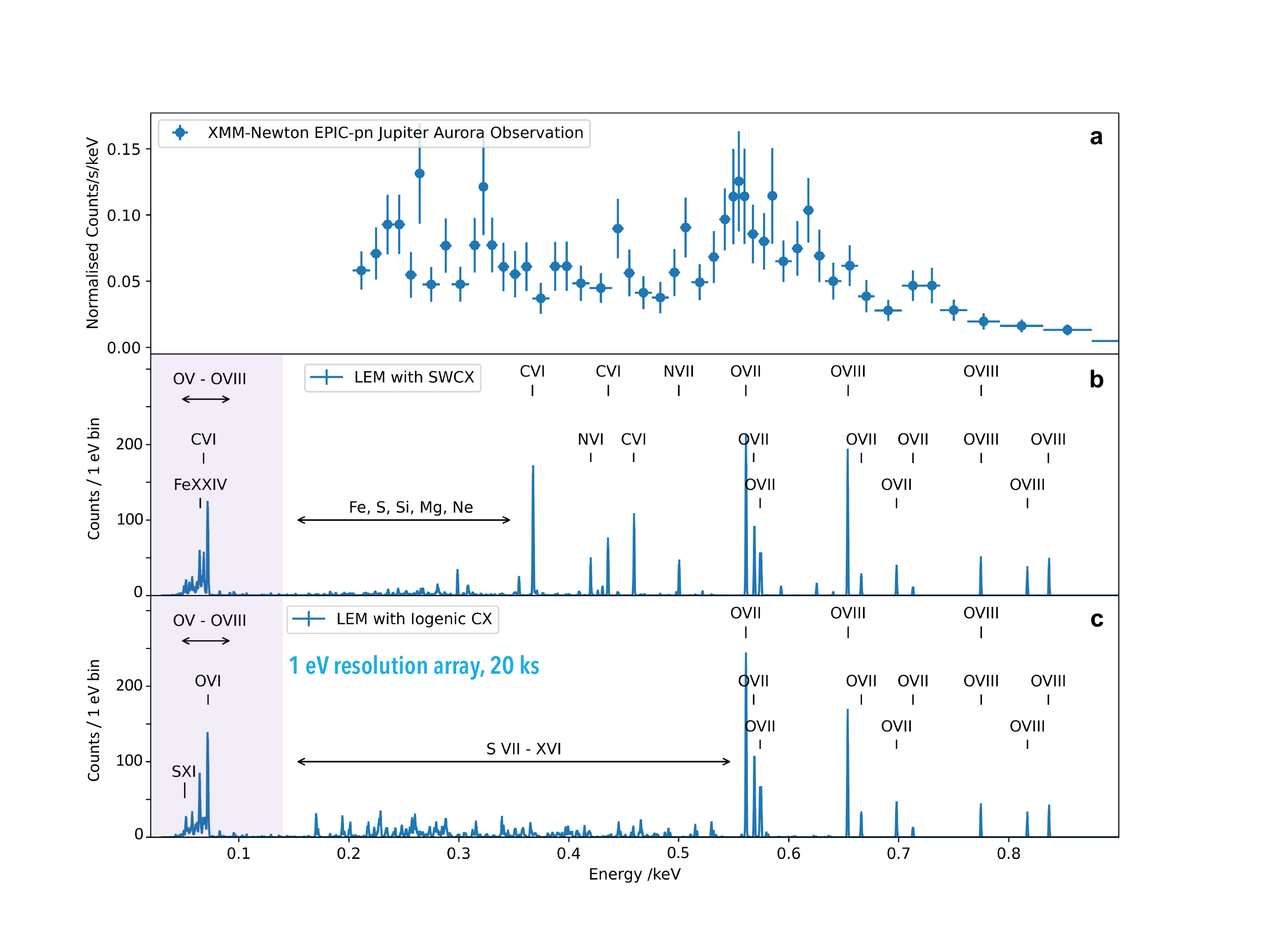}
}

\caption{\LEM\ will revolutionize solar system studies. ({\em a})
\XMM\ EPIC-pn spectrum of Jupiter's Northern X-ray aurora (130 ks). There
are two different models that fit the data well, but each offers
fundamentally different explanations for the processes that produce
Jupiter's aurora. The first model is a solar wind charge exchange model
(SWCX), implying entrance of solar wind into the Jovian system. The second
model is an Iogenic model, which uses charge exchange on the S and O ions
from Io's volcanos. ({\em b,c}) \LEM\ 1 eV resolution spectra for each
scenario. Counts per resolution element for a 20 ks exposure is shown;
background is not included for clarity (it is negligible because Jupiter is
under 1\am\ in size). The most prominent emission lines are marked. The
instrument may offer significant sensitivity at $E<0.1$ keV with the same
high spectral resolution, providing access to never-before-seen charge
exchange lines at low energies (an option currently under study; gray area
in the plots). \LEM\ will easily distinguish between these models and
address key questions on if and when Jupiter's magnetosphere is open to the
solar wind --- a point of longstanding debate in the heliophysics and
planetary science communities.}

\label{fig:jupiter}
\end{figure*}
%%%%%%%%%%%%%%%%%%%%%%%%%%%%%%%%%%%%%%%%%%%%%%%%%%%%%%%%%%%%%%%

%%%%%%%%%%
\smsk {\bi Hot ISM and X-ray binaries in nearby galaxies.}\/ In face-on
spirals, \lem\ will map the ISM dominated by emission lines and isolate the
X-ray binaries (XRB) spatially and using their time variability. In nearby
spirals such as M33 (Fig.\ \ref{fig:m33}), \lem\ will be able to study XRB
populations and star-forming regions that are generally not detected in the
Galaxy due to obscuration. We will be able to apply plasma diagnostics to
determine whether the gas in and around star-forming regions is in CIE, or
whether it has lost a significant amount of energy in expansion, while
remaining in a strongly overionized state. This determination will be
pivotal for understanding the formation of ISM superbubbles.

In accreting stellar-mass black hole systems, relativistically-broadened O-K
and Fe-L lines are fluoresced off the surface of the accretion disk. These
features can be used to constrain the accretion disk inner radius and/or
black hole spin.

%%%%%%%%%%
\smsk {\bi Jellyfish galaxies.}\/ \lem\ will map the process of ram-pressure
stripping of galaxies traveling through the gas in groups, clusters and the
Cosmic Web filaments, and of galaxies undergoing collisions with other
galaxies. This is an important mechanism by which the ICM and IGM are
enriched by metals. \lem\ will be particularly well-equipped to detect the
stripped gas and study its mixing with the ICM --- a process that strongly
depends on plasma physics and such properties as plasma viscosity, thermal
conduction and the structure of its magnetic fields.

Figure \ref{fig:jellyfish} illustrates how \lem\ will use its energy
resolution to separate the cool stripped gas, whose emission is dominated by
certain lines, from the hotter ICM, revealing the morphology of the stripped
tail, which strongly depends on plasma properties\cite{Roediger15}.
Furthermore, as stripping can bring neutral gas from the galaxy in direct
contact with the hot ICM, one can in principle expect the charge exchange
emission from their interface, with its tell-tale line signatures in the
\lem\ spectra. This is a completely unexplored area that holds great promise
for probing the ICM plasma physics. \lem\ will also study the effect of
galaxy collisions and ram pressure stripping on ISM and star formation.

%%%%%%%%%%
\smsk {\bi Shocks in outskirts of massive galaxies.}\/ As noted above
(\S\ref{sec:galform}), the emission from the outer regions of the CGM of
massive galaxies may be enhanced by resonant scattering of the CXB
photons\cite{Khabibullin19} and of the photons from the galaxy's own bright
central emission. This ``illumination'' in the resonant O{\sc vii}\/ line
may help \lem\ detect and map the outer reaches of the galactic halos, where
O{\sc vii} should be abundant but the intrinsic thermal emission is too
faint. It is at those virial regions that we expect to find shock fronts
from galaxy/galaxy mergers and matter infall. Recently, a radio phenomenon
was discovered that may be caused by those shocks (``odd radio circles,'' or
ORCs, around massive galaxies at several hundred kpc from their
centers\cite{Norris21}). \lem\ is well-equipped for detecting such shock
features via their associated density jumps in thermal plasma, highlighted
by the resonantly scattered light. This will uncover an additional piece of
the galaxy formation puzzle. Such measurements will have to be done at
$z>0.035$ in order to move the O{\sc vii}\/ resonant line from behind the MW
O{\sc vii}\/ triplet.

%%%%%%%%%%
\smsk {\bi Light echo on giant molecular clouds}\/ in the Galactic
Center. The SMBH in the center of our Galaxy is quiescent at present, but
may have been more active in past several hundred years. The light echo from
past explosions will be seen with a time delay in the fluorescent line
emission from the GC molecular clouds.\cite{Churazov17} While the bright Fe
fluorescent line at 6.4 keV is inaccessible for \lem, we will be able to
detect such light echo in the Si fluorescence. \lem\ will require only 4--5
pointings to map the entire region of the giant molecular clouds and map the
most recent history of activity of our own SMBH.

%%%%%%%%%%
\smsk {\bi Charge exchange emission}\/ in the Jovian magnetosphere, planets
and comets. \lem\ will offer revolutionary insights for heliophysics and
planetary science. One example is the Jovian system. Jupiter produces the
most powerful aurora in the solar system and these are key to understanding
which processes govern rapidly-rotating magnetospheres. It is unknown to
what extent the X-ray aurora are produced by the precipitation of material
that originates in Io's volcanos or whether they are indicative of the
direct entry of solar wind particles into the system\cite{Dunn22}.

Figure \ref{fig:jupiter} shows how \LEM's exquisite energy resolution will
finally solve this problem. The figure shows the unique new access LEM
provides to the low energy line emissions from planets. This will open an
entirely new paradigm, enabling the bright emissions in this spectral range
to be utilized to study previously inaccessible objects or to offer a step
change in capability to advance previous X-ray studies of Mars, Saturn,
comets, the Ice Giants and the heliosphere\cite{Bhardwaj07}.

A possibility to extend the \lem\ calibrated energy band down to $E=0.05$
keV is currently under study; while the effective area there would be small,
the CX signal at those energies is very luminous. This will also allow
diagnostics for species more closely linked to the standard UV lines of
C{\sc iv}, Si{\sc iv}, N{\sc v}, and O{\sc vi}. The sub-eV resolution of
\lem's central array would be critical in this very crowded spectral region.

%%%%%%%%%%
\smsk {\bi Earth magnetosphere.} A possibility to point \lem\ toward Earth
from an L1 orbit is currently under study. If it is possible, \lem\ will be
able to observe the SWCX emission from Earth's magnetopause and
magnetosheath, study their dynamics and their response to the varying solar
wind pressure --- a matter of longstanding interest for the heliophysics
community.

%%%%%%%%%%
In addition, the \lem\ observatory will greatly advance the studies of
variable black hole winds, flares in exoplanet host stars, protostar
accretion rates, interstellar dust in emission and absorption, and numerous
other fields. Given the big leap in observing capability that \lem\ will
bring, observers will most certainly come up with experiments that we cannot
even foresee at present.

\vspace*{-1mm}
\section{CONCLUSIONS}
\vspace*{-1mm}

The Line Emission Mapper Probe will fundamentally transform our
understanding of one of the most important questions in modern astrophysics
--- the formation of galaxies.  The unique ability of
this observatory to detect the low surface brightness X-ray emission from
the warm and hot gas in galaxy halos and in the filamentary structures
connecting larger mass structures, and to measure the temperature, entropy,
and elemental abundances will allow an understanding of the state of this
gas that is, at present, only a matter of theoretical speculation.
% The existing X-ray observatories such as
% \chandra\ and \xmm\ can access only the central, denser regions of this gas
% in galaxies and in cluster outskirts.
With its combination of wide field of view X-ray optic and energy-dispersive
microcalorimeter array, \LEM\ offers the opportunity for complete dynamical
and thermodynamic characterization of this gas out to a large fraction of
the virial radius in Milky Way-sized galaxies and beyond the virial radius
for galaxy clusters. \LEM\ will detect metals expelled by galaxies over
their lifetime and accumulated in the Cosmic Web filaments, and directly
measure the imprint of AGN feedback, stellar feedback, and merging over
cosmic time in galaxies halos and cluster filaments.

\LEM\ is the X-ray mission for the 2030s, and it should be the first of the
exciting new line of astrophysics Probes.  The primary science of \LEM\ is
of great interest to the entire astrophysical community, not just the X-ray
community. Most of the \LEM\ observing time will be dedicated to the guest
observer program.  The combination of wide field of view and 1--2 eV
spectral resolution of the calorimeter offers a unique host of guest observer
investigations covering every area of X-ray astrophysics, including studies of
AGN, stellar astrophysics, exoplanets, Solar system physics, and time domain
and multi-messenger astrophysics.  \LEM\ also has significant synergies with
virtually every ground-based and space-based observatory planned to be in
operation in the 2030s, including \SKA, Roman Space Telescope, CMB-S4, and
many others.  Finally, the technologies required to implement this mission
--- the lightweight high-resolution X-ray optics and the large
microcalorimeter arrays --- are rapidly maturing.  With small, well-defined,
ongoing investments in technology development, \LEM\ is already on the path
to meet its key mission milestones.  \LEM\ offers the astrophysics community
paradigm-changing science in a Probe package.

%\section*{REFERENCES}

%%%%%%%%%%%%%%%%%%%%%%% REFERENCES:
\small

\end{document}